\documentclass[prd,twocolumn,superscriptaddress,altaffilletter,amssymb,showpacs,nofootinbib]{revtex4}
\usepackage[dvips]{graphicx}
\usepackage{amsmath}

\begin{document}
\title{Flat Central Density Profiles from Scalar Field Dark Matter Halos}

\author{Argelia Bernal}\email{abernal@fis.cinvestav.mx}
\author{Tonatiuh Matos\footnote{http://www.fis.cinvestav.mx/$\sim$tmatos}}\email{tmatos@fis.cinvestav.mx}
\affiliation{Departamento de Fisica, Centro de Investigaci\'{o}n y
de Estudios Avanzados del IPN, AP 14-740, 07000 D.F., M\'{e}xico}
\affiliation{Part of the Instituto Avanzado de Cosmolog\'ia (IAC)
collaboration http://www.iac.edu.mx/}

\author{Dar\'{\i}o N\'u\~{n}ez}\email{nunez@nuclecu.unam.mx}
\affiliation{Instituto de Ciencias Nucleares, Universidad Nacional
Aut\'onoma de M\'{e}xico, AP 70-543, 04510 D.F., M\'{e}xico.}
\affiliation{On Sabbathicak leave at Max-Planck-Institut f\"ur
Gravitationsphysik, Albert Einstein Institut, 14476 Golm,
Germany.}
 \affiliation{Part of the Instituto Avanzado de
Cosmolog\'ia (IAC) collaboration http://www.iac.edu.mx/}

\date{\today}
\begin{abstract}
Scalar fields endowed with a cosh potential behaves in the linear
regime, exactly as the cold dark matter (CDM) model. Thus, the
scalar field dark matter (SFDM) hypothesis predicts the same
structure formation as the CDM model. The free parameters of the
SFDM model are determined by cosmological observations. In a
previous work we showed that if we use such parameters, the scalar
field collapses forming stable objects with a mass around
$10^{12}M_{\odot}$. In the present work we use analytical
solutions of the flat and weak field limit of the Einstein-
Klein-Gordon equations and show that the SFDM density profile
corresponds to a halo with an almost flat central density and that
it coincides with the CDM model in a broad outer region. This
result could solve the problem of the density cusp DM halo in
galaxies without any additional hypothesis, supporting the
viability of the SFDM model.
\end{abstract}

\maketitle

\section{Introduction}
\label{sect:int}

The Lambda Cold Dark Matter ($\Lambda $CDM) model has recently
shown an enormous predictive power. It can explain the structure
formation of the Universe, its accelerated expansion, the micro
Kelvin fluctuation of the Cosmic Microwave Background Radiation,
etc. Nevertheless, some issues around this model related to
the formation of galaxies have arisen since the time it was
originally proposed and remain to date. The CDM paradigm
predicts a density profile which corresponds to the
Navarro-Frenk-White (NFW) profile, Navarro et al, 1997, given by
\begin{equation}
\rho _{NFW}=\frac{\rho _{0}}{\frac{r}{r_0}(\frac{r}{r_0}+1)^{2}}.
\label{NFW}
\end{equation}

\noindent However, this profile seems to have some differences
to the observed profiles of LSB galaxies. In this work we show
that a flat central profile naturally arises within the scalar
field dark matter hypothesis, implying that the central region of
galaxies can distinguish between CDM and SFDM.

We work within the specific context of the so--called `strong,
self--interacting scalar field dark matter' (SFDM) hypothesis that
has been developed by several authors, Guzm\'{a}n  and Matos 2000,
Matos and Ure\~{n}a-L\'{o}pez 2000, 2001; Ure{\~n}a-L{\'o}pez,
Matos and Becerril 2002; Matos and Guzm\'{a}n 2001; Alcubierre et
al. 2002, 2003; Ure{\~n}a-L{\'o}pez 2002; B\"ohmer and Harko 2007;
(see also Peebles 2000). { A first proposal of the SFDM hypothesis
appeared in a couple of papers by Ji and Sin 1994. They took a
massive scalar field and were able to fit observations comming
from some galaxies, taken into account also the contribution of
baryons. A key point in this work was the use of the so-called
excited configurations in which the radial profile of the scalar
fields has nodes. From that, they determined that the mass of the
scalar field should be of order of $\sim 10^{-24}$ eV.

A next proposal appeared in a paper by Schunck 1998. He shows that
a massless complex scalar field can be used as a dark matter model
in galaxies to fit the rotation curves. In this model the internal
frequency of the field plays the role of an adjustable parameter,
and the radial profile of the scalar field also has nodes.

However, as pointed out in Guzm\'an and Ure\~na-L\'opez 2005 (see
also Guzm\'an and Ure\~na-L\'opez 2004), this last proposal cannot
be realistic because a massless scalar field (whether real or
complex as in Schunck 1998) cannot form a gravitationally bound
configuration (see Seidel and Suen 1994).}

The key idea of the SFDM scenario is that the dark matter
responsible for structure formation in the Universe is a real
scalar field, $\Phi $, minimally coupled to Einstein gravity with
self-interaction parametrized by a potential energy of the form
(see also Sahni and Wang 2000)
\begin{equation}
V(\Phi )=V_{0}\left[\cosh(\lambda \sqrt{\kappa_{0}}\Phi )-1\right],
\label{coshpot}
\end{equation}
\noindent where $V_{0}$ and $\lambda $ are the only two free
parameters of the model, $\kappa _{0}=8\pi G$ and we employ
natural units $\hbar=c=1$. The effective mass of the
scalar field is given by $m_{\Phi}^{2}=\kappa _{0}V_{0}\lambda^{2}$.
\newline

The advantage of the SFDM model is that it is insensitive to
initial conditions and the scalar field behaves as CDM once it
begins to oscillate around the minimum of its potential. In this
case, it can be shown (see Matos and Ure\~{n}a-L\'{o}pez 2000,
2001) that the SFDM model is able to reproduce all the successes
of the standard $\Lambda $CDM model above galactic scales.

Furthermore, it predicts a sharp cut-off in the mass power
spectrum due to its quadratic nature, thus explaining the observed
dearth of dwarf galaxies, in contrast with the possible excess
predicted by high resolution N-body simulations with standard CDM,
see Matos and Ure\~{n}a-L\'{o}pez 2001.

The best--fit model to the cosmological data can be deduced from
the current densities of dark matter and radiation in the Universe
and from the cut--off in the mass power spectrum that constrains
the number of dwarf galaxies in clusters. The favored values for
the two free parameters of the scalar field potential are found to
be, Matos and Ure\~{n}a-L\'{o}pez 2001:
\begin{eqnarray}
\lambda &\simeq &20,  \nonumber \\
V_{0} &\simeq &(3\times 10^{-27}\,m_{Pl})^{4}\,,  \label{V0}
\end{eqnarray}
\noindent where $m_{Pl}\equiv G^{-1/2}\approx 10^{-5}$g is the
Planck mass.

This implies that the effective mass of the scalar field should be
$m_{\Phi}\simeq 9.1\times 10^{-52}\,m_{Pl}=1.1\times 10^{-23}$
eV.\newline

Let us explain why we suspect that the scalar field could be the
dark matter at galactic scales as well. There are three main
reasons.

The first reason is that numerical simulations suggest that the
critical mass for the case considered here, using the scalar
potential (\ref{coshpot}), and the parameters given by
Eq.~(\ref{V0}), is approximately, Alcubierre et. al. 2002
\begin{equation}
M_{crit}\simeq
0.1\frac{m_{Pl}^{2}}{\sqrt{\kappa_{0}V_{0}}}=2.5\times
10^{13}M_{\odot }\,.  \label{masa1}
\end{equation}

This was a surprising result. The critical mass of the model shown
in Matos and Ure\~{n}a-L\'{o}pez 2000, 2001, is of the same order
of magnitude of the dark matter content of a standard galactic
halo. Observe that the parameters of the model, Eq.~(\ref{V0}),
were fixed using cosmological observations. The surprising result
consisted in the fact that using the same scalar field for
explaining the dark matter at cosmological scales, it will always
collapse with a preferred mass which corresponds to the halo of a
real galaxy. Thus, this result is a prediction of the cosmological
SFDM model for galaxy formation.

The second reason is that during the linear regime of cosmological
fluctuations, the scalar field and a dust fluid, like CDM, behave
in the same way. The density contrast in CDM and in the SFDM
models evolve in exactly the same form and then both models
predict the same large scale structure formation in the Universe
(see Matos and Ure\~{n}a-L\'{o}pez 2001). The differences between
the CDM and SFDM models begin to appear in the non linear regime
of structure formation, so that there will be differences in their
predictions on galaxy formation.

The third reason is the topic of this work. A scalar field object
(e.g. an oscillaton) contains a flat central density profile, as
seems to be the case in galaxies.

In the case of the SFDM, the strong self-interaction of the scalar
field results in the formation of solitonic objects called
`oscillatons', which have a mass of the order of a galaxy (see for
example Ure\~{n}a-L\'{o}pez 2002, Ure\~{n}a-L\'{o}pez et al 2002
and Alcubierre et al 2003. Also Seidel and Suen 1991, 1994, Hawley
and Choptuik 2000, and Honda and Choptuik 2001). In this work we
will show that these models contain an almost flat central density
profile, ${\it i.e.}$, they do not exhibit the cusp density
profiles characteristic of the standard CDM hypothesis.

Before starting with the description, we want to emphazise the
fact that the scalar field has no interaction with the rest of the
matter, thus, it does not follow the standard lines of reasoning
for the particle-like candidates for dark matter. The scalar field
was not thermalized, that is, the scalar field forms a Bose
condensate, and thus behaves strictly as cold dark matter from the
beginning.

The rest of the paper is organized as follows. In the next
section we use the fact that Galaxies have a weak gravitational
field and thus their space-time is almost flat. The main goal of
this section is to study the physics provoking the flatness
behavior of the density profiles at the center of the oscillatons.
Some results of this section intersect with those presented by
Jin and Sin 1994, where they studied the behavior of the weak
field limit of a complex scalar field. We remark that we do our
analysis for a real scalar field and some differences do arise due
to a different current conservation. In section \ref{sect:wf} we
present our contribution making an analysis of the
Einstein-Klein-Gordon (EKG) equations in the relativistic weak
field limit and solve them for the perturbed metric coefficients.
Then, we compare these solutions with the ones obtained by solving
numerically the complete EKG system and using the whole potential
(\ref{coshpot}), we show that the relativistic weak field limit is
indeed a very good approximation. In section \ref{sect:sf} we
compute the energy density of the scalar field obtained in section
\ref{sect:wf}, and compare it with actual observations of LSB
galaxies from which the density is inferred from the rotational
curves, showing a good matching in the external regions and a
match at least similar in some of the internal regions, in any
case, better than a fit with a cusp-like behavior of the density.
Finally in section \ref{sect:conc} we give our conclusions.

\section{Physics of the Scalar Field. Flat space-time case}
\label{sect:flat}

In this section we derive the physics of the scalar field in an
analogous way as it was done for the complex scalar field by Jin and
Sin 1994 (see also Lee and Koh 1996). In a normal dust collapse,
as for example in CDM, there is in principle nothing to avoid that
the dust matter collapses all the time. There is only a radial
gravitational force that provokes the collapse, and to stop it,
one needs to invoke some virialization phenomenon. In the scalar
field paradigm the collapse is different. The energy momentum
tensor of the scalar field is
\begin{equation}
T_{\mu \nu }=\Phi _{,\mu }\Phi _{,\nu }-\frac{g_{\mu \nu }}{2}\,
\left[ \Phi^{,\alpha }\Phi _{,\alpha }+2V(\Phi )\right] \,.
\label{eq:set}
\end{equation}
We will consider spherical symmetry, and work with the line element
\begin{equation}
ds^{2}=-e^{2\nu }dt^{2}+e^{2\mu }dr^{2}+r^{2}d\Omega ^{2}\,,
\label{elem}
\end{equation}
with $\mu =\mu (r,t)$ and $\nu =\nu (r,t)$, being this last
function the Newtonian potential. The energy momentum tensor of
the scalar field has then the components
\begin{eqnarray}
&&-{T^{0}}_{0} =\rho _{\Phi }= \frac{1}{2} \left[ e^{-2\nu
}{\dot{\Phi}}^{2}+e^{-2\mu }\Phi^{\prime 2}+
2V(\Phi )\right]  \label{tensor1} \\
&&T_{01} ={\cal P}_{\Phi }=\dot{\Phi}\Phi^{\prime }  \label{tensor2} \\
&&{T^{1}}_{1} =p_{r}=\frac{1}{2}\left[ e^{-2\nu }{\dot{\Phi}}^{2}+
e^{-2\mu}\Phi ^{\prime 2}-2V(\Phi )\right]  \label{tensor3} \\
&&{T^{2}}_{2} =p_{\bot }=\frac{1}{2} \left[ e^{-2\nu
}{\dot{\Phi}}^{2}- e^{-2\mu }\Phi ^{\prime 2}- 2V(\Phi )\right]
\label{tensor4}\nonumber\\
\end{eqnarray}
and also ${T^{3}}_{3}={T^{2}}_{2}$. These different components are
identified as the energy density $\rho _{\Phi }$, the momentum
density ${\cal P}_{\Phi }$, the radial pressure $p_{r}$ and the
angular pressure $p_{\bot }$. The integrated mass is defined by
\begin{equation}
M(x)=4\pi \int_{0}^{x}\rho _{\Phi }(X)\,X^{2}dX\,. \label{mass}
\end{equation}

The radial and angular pressures are two natural components of the
scalar field which stop the collapse, avoiding the cusp density
profiles in the centers of the collapsed objects. This is the main
difference between the normal dust collapse and the SFDM one. The
pressures play an important roll in the SFDM equilibrium. In order
to see this, and considering that galaxies are almost
flat, we conclude that the Newtonian approximation should be sufficient to
describe the processes. In this section we will take the flat space-time
approximation.

Thus, we study a massive oscillaton without self-interaction
($i.e.$ with potential $V=\frac{1}{2}m_{\Phi }^{2}\Phi ^{2}$), in
the Minkowski background ($\mu\sim\nu \sim 0$). Even though it is
not a solution to the Einstein equations as we are neglecting the
gravitational force provoked by the scalar field, the solution is
analytic and it helps us to understand some features that appear
in the non-flat oscillatons.\newline

In a spherically symmetric space-time, the Klein-Gordon equation
$\eta^{\alpha\beta}\partial_{\alpha}\partial_{\beta}\,\Phi-dV/d\Phi
=0$, (where $\eta^{\alpha\beta}\partial_{\alpha}\partial_{\beta}$
stands for the D'Alambertian), reads
\begin{equation}
\Phi^{\prime \prime }+\frac{2}{r}\Phi^{\prime }- m_{\Phi}^{2}\Phi
=\ddot{\Phi}, \label{eq:KGp}
\end{equation}
\noindent where over-dot denotes $\partial /\partial t$ and prime
denotes $\partial /\partial r$. The exact general solution for the
scalar field $\Phi$ is
\begin{equation}
\Phi(t,r) = \frac{e^{\pm ikr}}{r} e^{\pm i\omega t}, \label{eq:flatphi}
\end{equation}
\noindent and we obtain the dispersion relation
$k^{2}=\omega^{2}-m_{\Phi}^{2}$. For $\omega >m_{\Phi }$ the
solution is non-singular and vanishes at infinity. We will
restrict ourselves to this case. It is more convenient to use
trigonometric functions and to write the particular solution in
the form
\begin{equation}
\Phi (t,x)=\Phi _{0}\frac{\sin (x)}{x}\cos (\omega t), \label{fphi}
\end{equation}
\noindent where $x=kr$. It oscillates in harmonic manner in time.
The scalar field can be considered to be confined to a finite
region, see Ure\~{n}a-L\'{o}pez 2002, and Ure\~{n}a-L\'{o}pez et
al 2002.
\begin{figure}[htp]
\begin{center}
\includegraphics[width=8.0cm]{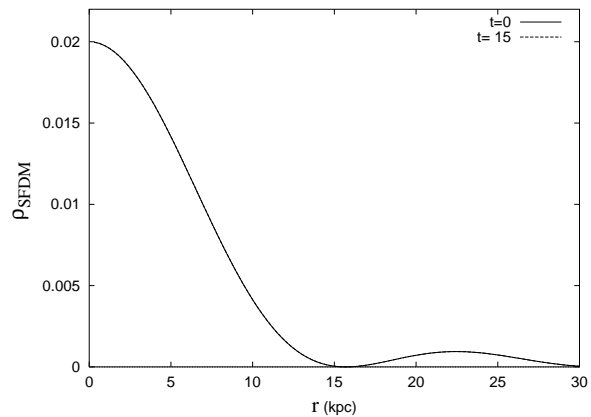}
\caption{The energy density profile for the Scalar Field Dark
Matter model (see also Schunck 1997). The parameters used in this
plot are $m_{\Phi}=20,\protect\omega=20.001,\Phi_{0}^{2}=5\
10^{-2}$. The density is given in arbitrary units and $r$ is given
in kpc.} \label{fig:SFDM}
\end{center}
\end{figure}
The analytic expression for the scalar field energy density
derived from Eq. (\ref{fphi}) is
\begin{eqnarray}
\rho _{\Phi}=&&\frac{{\Phi_0}^2\,k^2}{2\,x^2}\left[\left(\left(\frac{\sin(x)}{x}-\cos(x)\right)^2 \right. \right.  \\  && \left. \left. -k^2\,\sin^2(x)\right)\,\cos^2(\omega\,t)+ \omega^2\,\sin^2(x)\right], \nonumber
\label{rho}
\end{eqnarray}
\noindent which oscillates with a frequency $2\omega t$. Observe
that close to the central regions of the object, the density of
the oscillaton behaves like
\begin{equation}
\rho_{\Phi }\sim \frac{1}{2}{\Phi_{0}}^{2}k^{2} \left[
\omega^{2}-k^{2}\cos^{2}\left( \omega \,t\right) \right] +O(x^{2}),
\end{equation}
which implies that when $x\rightarrow 0$ the central density
oscillates around a fixed value.

On the other hand, the asymptotic behavior when $x\rightarrow
\infty $, is such that $\rho _{\Phi }\sim 1/x^{2} $, $i.e.$ far
away from the center, in this approximation, the flat oscillaton
density profile behaves like the isothermal one. The mass function
oscillates around $M\sim x$, as usual for the galactic halos.

In order to understand what is happening within the object,
observe that the KG equation can be rewritten in a more convenient
form in terms of the energy density, as
\begin{equation}
\frac{\partial \rho _{\Phi }}{\partial t}-
\frac{1}{r^{2}}\frac{\partial }{\partial r} \left( r^{2}{\cal
P}_{\Phi }\right) =0. \label{kgflat}
\end{equation}

This last equation has a clear interpretation: Since its form
looks like the conservation equation, $\dot{\rho}+\nabla \cdot
\vec{J}=0 $, equation (\ref{kgflat}) represents the conservation
of the scalar field energy. It also tells us that there is a
scalar field current given by
\begin{eqnarray*}
\vec{J}_{\Phi } &=&-{\cal P}_{\Phi }\vec{r} \\
&=&\Phi_{0}^{2}\frac{k\omega }{2}\left[ x\cos(x)-\sin(x)\right]
\frac{\sin(x)\sin(2\omega t)}{x^{3}}\vec{r}.
\end{eqnarray*}

Observe that the quantity involved in this current is the scalar
field momentum density (\ref{tensor2}). Although the flux of
scalar radiation at large distances does not vanish, there is not
a net flux of energy, as it can be seen by averaging the scalar
current on a period of a scalar oscillation. We also see that the
only transformation process is that of the scalar field energy
density into the momentum density, and viceversa. For the
realistic values (\ref{V0}) this transfer is very small.

\begin{figure}[htp]
\begin{center}
\includegraphics[width=8.0cm]{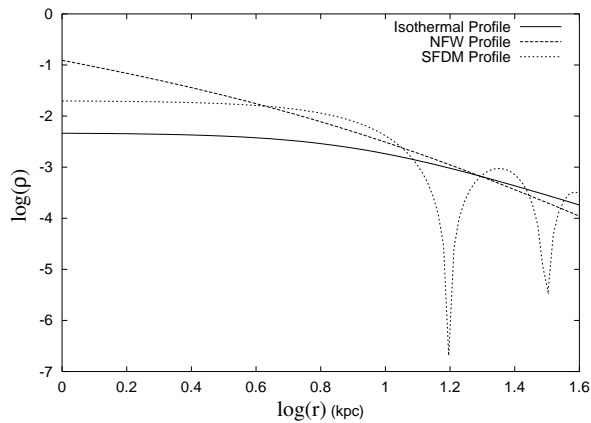}
\caption{Comparison between the energy density profile for the
Scalar Field Dark Matter model with the NFW and the Isothermal
models. The parameters for the isothermal model are $\protect\rho
_{Iso}=0.3/(r^{2}+8^{2})$ and for the NFW profile are
$\protect\rho _{NFW}=10/(r(r+8)^{2})$. The parameter used for the
SFDM model are the same as in the previous figure.}
\label{fig:compar}
\end{center}
\end{figure}

In Fig. \ref{fig:SFDM} we show the behavior of the SFDM density
profile for a typical galaxy and in Fig. \ref{fig:compar} we show
the comparison between the NFW, the isothermal and the SFDM
density profiles for the same galaxy. Observe that the SFDM and
NFW profiles remain very similar up to 10 kpc, then the SFDM
profile starts to follow the isothermal one.

In the next section we will see that if the gravitational force is
taken into account, the oscillaton is more confined (see also
Ure\~{n}a-L\'{o}pez 2002, Ure\~{n}a-L\'{o}pez et al 2002, and
Alcubierre et al 2003). The parameters used in the figures,
correspond to a middle size galaxy.\newline

\section{Weak Field Limit Equations}
\label{sect:wf}

{In this section we derive a novel method for integrating the
perturbed EKG equations and show that the solutions are in very
good agreement with the numerical ones. This allow us to use these
solutions to fit the rotation curves of several observed LSB
galaxies.}

Within general relativity, the evolution of the scalar and
gravitational fields are governed by the coupled EKG equations,
the last one appearing from the conservation of the energy-momentum
tensor
\begin{eqnarray}
R_{\alpha\beta}&=&\kappa_0S_{\alpha\beta},\\
T^{\alpha \beta};_{\beta}&=&\Phi^{,\alpha}(\nabla^2 - m^2)\Phi=0,\label{conservation}
\end{eqnarray}
here $R_{\alpha\beta}$ is the Ricci tensor,
$S_{\alpha\beta}=T_{\alpha\beta}-
(1/2)g_{\alpha\beta}T^{\lambda}_{\phantom{\lambda}\lambda}$,
and $\nabla^2=
(1/\sqrt{-g})\partial_{\mu}[\sqrt{-g}g^{\mu\nu}\partial_{\nu}]$ is
the covariant D'Alambertian operator.

For simplicity, we continue to consider the non-static spherically
symmetric case, given by Eq.~(\ref{elem}). As usual in the weak
field limit, we suppose the metric to be close to the Minkowski
metric $\eta_{\alpha\beta}$
\begin{equation}
g_{\alpha\beta}=\eta_{\alpha\beta}+h_{\alpha\beta},
\end{equation}
where $|h_{\alpha\beta}|\ll 1$, then we will consider an expansion
of the functions in the metric of the form
\begin {eqnarray}\label{perturbations}
e^{2\nu(t,r)}=1+\epsilon^{2}2V(t,r)+O(\epsilon^{4}),\nonumber\\
e^{2\mu(t,r)}=1+\epsilon^{2}2U(t,r)+O(\epsilon^{4}),
\end {eqnarray}
where $\epsilon $ is an expansion parameter. We also consider that
the spatial and time derivatives of the geometric quantities are
regarded like
\begin{equation}\label{partials}
\frac{\partial}{\partial t}\sim \frac{\partial }{\partial r}.
\end{equation}
then to first order in $\epsilon^{2}$ the Ricci tensor components
are respectively
\begin{eqnarray}\label{riccitensor}
R_{tt}&=&[V,_{rr}-U,_{tt}+\frac{2}{r}V,_{r}],\nonumber\\
R_{rr}&=&[U,_{tt}-V,_{rr}+\frac{2}{r}U,_{r}],\nonumber\\
R_{tr}&=&[\frac{2}{r}U,_{t}].
\end{eqnarray}

On the other hand, the source is computed in the flat space in
this case as well, thus the scalar field satisfies to
$\epsilon^{2}$ order the Eq.~(\ref{eq:KGp}) as in the previous
section.

It is important to emphasize that the relation (\ref{partials})
is the lowest one in the geometric fields, but it does not consider
small velocities for the sources. This is different from the
Newtonian limit where the derivative relation for the scalar field
is $\partial_r\sim \epsilon \partial_t$ and $\partial_t\sim
\epsilon \partial_r$ for the geometric fields (see Seidel \& Suen
1990, and Guzm{\'a}n \& Ure{\~n}a-L{\'o}pez 2004).

Consistent with the $ T_{\alpha\beta}$ computed in the flat
space-time, the right hand side  elements in Einstein's equations,
are written as
\begin{eqnarray}\label{emtensor}
S_{tt}&=&\Phi,_t\Phi,_{t}-\frac{1}{2}m^2|\Phi|^2,\nonumber\\
S_{rr}&=&\Phi,_r\Phi,_{r}+\frac{1}{2}m^2|\Phi|^2,\nonumber\\
S_{tr}&=&\Phi,_r\Phi,_{t}.
\end {eqnarray}
In this case it is convenient to introduce the dimensionless
quantities
\begin {equation}
x=mr,\quad \tau=mt,\quad \Omega=\frac{\omega}{m},
\end {equation}
where we note that the bosonic mass $m$ is the natural scale for
time and distance. In terms of these new variables the general
solution to Eq.~(\ref{eq:flatphi}) takes the form
\begin{equation}
\Phi(\tau,x)=\frac{1}{x}\exp(\pm i x\sqrt{\Omega^2-1})\exp(\pm i\Omega \tau).
\end{equation}
The physical properties of the solution depend on the ratio
$\Omega\equiv\omega/m$. For $\Omega < 1$ the solution decays
exponentially but it is singular at $r=0$. On the other hand,
$\Omega > 1 $ allows for non-singular solutions which  vanish at
infinity. We will restrict ourselves to this case. We will
write the particular solution in the form
\begin{equation}\label{scalarf}
\sqrt{\kappa_0}\Phi(\tau,x)=\phi(x)\cos(\Omega\tau),
\end{equation}
where the spatial function is given by
\begin{equation}\label{spatialscalarf}
\phi(x)=\phi_0\frac{\sin(x\sqrt{\Omega^2-1})}{x}.
\end{equation}

Because of the functional form of the scalar field (\ref{scalarf})
we introduce the following {\it ansatz} for the metric
perturbations (\ref{perturbations})
\begin{eqnarray}
V(\tau,x)&=&V_0(x)+V_2(x)\cos(2\Omega\tau),\nonumber\\
U(\tau,x)&=&U_0(x)+U_2(x)\cos(2\Omega\tau). \label{eq:VU}
\end{eqnarray}
We adopt the following method to solve the EKG equations. In the
first approximation we substitute the solution of a lower
approximation (the flat case) into the next approximation (with
$\epsilon^2$) and solve the resulting differential equations. As
we will see, this standard approximation works well in our case.
In terms of these expressions the Einstein's equations
$R_{\alpha\beta}=\kappa_0S_{\alpha\beta}$ finally read
\begin{eqnarray}\label{Einsteine}
\epsilon^2\left[\frac{4}{x}U_2 \right]&=&\frac{1}{2}\phi\phi,_x,\\
\epsilon^2\left[V_0,_{xx}+\frac{2}{x}V_0,_x \right]&=&\frac{1}{2}\left(\Omega^2-
\frac{1}{2}\right)\phi^2,\nonumber\\
\epsilon^2\left[V_2,_{xx}+\frac{2}{x}V_2,_x+4\Omega^2U_2 \right]&=&-\frac{1}{2}
\left(\Omega^2+\frac{1}{2}\right)\phi^2,\nonumber\\
\epsilon^2\left[-V_0,_{xx}+\frac{2}{x}U_0,_x\right]&=&\frac{1}{2}\left(\phi,_x^2+
\frac{1}{2}\phi^2\right),\nonumber\\
\epsilon^2\left[-V_2,_{xx}+\frac{2}{x}U_2,_x-4\Omega^2U_2\right]&=&\frac{1}{2}
\left(\phi,_x^2+\frac{1}{2}\phi^2\right). \nonumber
\end{eqnarray}

\subsection{Scaling properties}

From system (\ref{Einsteine}) we know that the scalar field's
maximum amplitude $\phi(0)=\phi_0\sqrt{\Omega^2-1}$ could be taken
as the expansion parameter $\epsilon$ and in this case $\Omega$
must be of order $1$. Then it is always possible to solve the
system (\ref{Einsteine}) ignoring $\epsilon$ and replacing $\phi$
by its normalized function
\begin {equation}\label{Phin}
\hat{\phi}(x)=\frac{\sin(x\sqrt{\Omega^2-1})}{x\sqrt{\Omega^2-1}}.
\end {equation}
Solutions $\hat{\phi}$, $U_0$, $U_2$, $V_0$, $V_2$ of this
normalized system depend only of the arbitrary characteristic
frequency $\Omega$ which modulate the wave length of $\hat{\phi}$.
On the other hand, for each value of $\Omega$ there is a complete
family of solutions of the scalar field $\phi$ and the metric
perturbations $h_{\alpha\beta}$
 which are related to each other by
a scaling transformation characterized by $\phi_0$
\begin{eqnarray}\label{sols}
\sqrt{\kappa_0}\Phi&=&\phi_0\sqrt{\Omega^2-1}\hat{\phi}\cos(\Omega\tau),
\\ h_{rr}&=&\phi_0^2(\Omega^2-1)[2U_0+2U_2\cos(2\Omega\tau)],
\nonumber\\
h_{tt}&=&-\phi_0^2(\Omega^2-1)[2V_0+2V_2\cos(2\Omega\tau)],\nonumber
\end{eqnarray}
In this context the weak field limit condition $h_{\alpha\beta}\ll 1$
translates into
\begin{equation}\label{Weakfieldc}
 \phi_0^2(\Omega^2-1)|2V|\ll 1,\quad  \& \quad \phi_0^2(\Omega^2-1)|2U|\ll 1.
\end{equation}
Here we will introduce a specific notation for the spatial
functions of the metric perturbations:
\begin{eqnarray}\label{eqs:hs}
h_{rr}^{(0)}&=&\phi_0^2(\Omega^2-1)2U_0,\\
h_{rr}^{(2)}&=&\phi_0^2(\Omega^2-1)2U_2, \nonumber \\
h_{tt}^{(0)}&=&\phi_0^2(\Omega^2-1)2V_0, \nonumber \\
h_{tt}^{(2)}&=&\phi_0^2(\Omega^2-1)2V_2. \nonumber
\end{eqnarray}

\subsection{Metric perturbations solutions}

The system of equations Eqs.~(\ref{Einsteine}) can be solved and
the spatial functions of the metric perturbations have analytic
solutions given by
\begin{eqnarray}\label{Einsteins}
\lefteqn{U_2=\frac{1}{8(\Omega^2-1)}\Big[-\frac{\sin^2(x\sqrt{\Omega^2-1})}
{x^2}+{}}\nonumber\\
& & {}+\frac{\sqrt{\Omega^2-1}}{2}\frac{\sin(2x\sqrt{\Omega^2-1})}{x}\Big],
\nonumber\\
\lefteqn{V_0=\frac{(2\Omega^2-1)}{8(\Omega^2-1)}\Big[
\frac{\sin(2x\sqrt{\Omega^2-1})}{2x\sqrt{\Omega^2-1}}-{}}\nonumber\\
& & {}-Ci(2x\sqrt{\Omega^2-1})+{}\nonumber\\
& & {}+\ln(2x\sqrt{\Omega^2-1})\Big] -\frac{C_{V01}}{x}+C_{V02},\nonumber\\
\lefteqn{V_2=\frac{1}{8(\Omega^2-1)}\Big[ \frac{\sqrt{\Omega^2-1}}{2}
\frac{\sin(2x\sqrt{\Omega^2-1})}{x}+{}}\nonumber\\
& & {}+Ci(2x\sqrt{\Omega^2-1})-{}\nonumber\\
& & {}-\ln(2x\sqrt{\Omega^2-1})\Big]+C_{V22},\nonumber\\
\lefteqn{U_0=\frac{1}{8(\Omega^2-1)} \Big[-\frac{1}{2x^2}+\frac{1}{2}
\frac{\cos(2x\sqrt{\Omega^2-1})}{x^2}-{}}\nonumber\\
& & {}-\frac{1}{2}\frac{\sin(2x\sqrt{\Omega^2-1})}{x\sqrt{\Omega^2-1}}
\Big]+\frac{C_{V01}}{x}+C_{U01},
\end{eqnarray}
where $Ci$ is the cosine integral function and $C_{V01}$, $C_{V02}$, $C_{V22}$,
and $C_{U01}$ are integration constants.

\subsection{Weak Field Validity Range}\label{WFVR}

From equation (\ref{eq:flatphi}) it is evident that, in the limit
in which we are working with, the KG equation is decoupled from
Einstein equations. Imposing regularity at the origin and
asymptotic flatness to the KG solution we have chosen
(\ref{scalarf}) with (\ref{spatialscalarf}) as our scalar field
particular solution where $\phi_0$ and $\Omega > 1$ are still free
parameters.

On the other hand, regularity at $x=0$ requires
$h_{rr}(x=0,\tau)=0$ which implies
\begin{equation}\label{Constantes}
C_{V01}=0, \quad C_{U01}=\frac{\Omega^2}{8(\Omega^2-1)},
\end{equation}
then for the perturbations $C_{V02}$ and $C_{V22}$ are still free
integration constants. Now we will describe the asymptotic
behavior of these perturbations. Due to $U_2$ being at least one order
of magnitude smaller than $U_0$ and its value oscillating around
zero, it is $U_0$ which determines the behavior of $h_{rr}$. The
$U_0$ value starts to oscillate, very near to the origin,  around
$C_{U01}$ keeping this behavior asymptotically. Then the
asymptotic value of $h_{rr}$ is the finite $C_{U01}$ value.
Contrary to this $h_{tt}$, due to the logarithm terms in $V_0$ and
$V_2$, $h_{tt}$, is singular at infinity. Thus, the weak field
condition (\ref{Weakfieldc}) is fulfilled only in a finite spatial
region around the origin, $i.e.$, due the the approximation the
solution is contained in a box, for which the walls are sufficiently far
away from the center of the solution. We will say that this is the
region where our weak field approximation is valid.

\begin{figure}[htp]
\begin{center}
\includegraphics[width=6.5cm, angle=270]{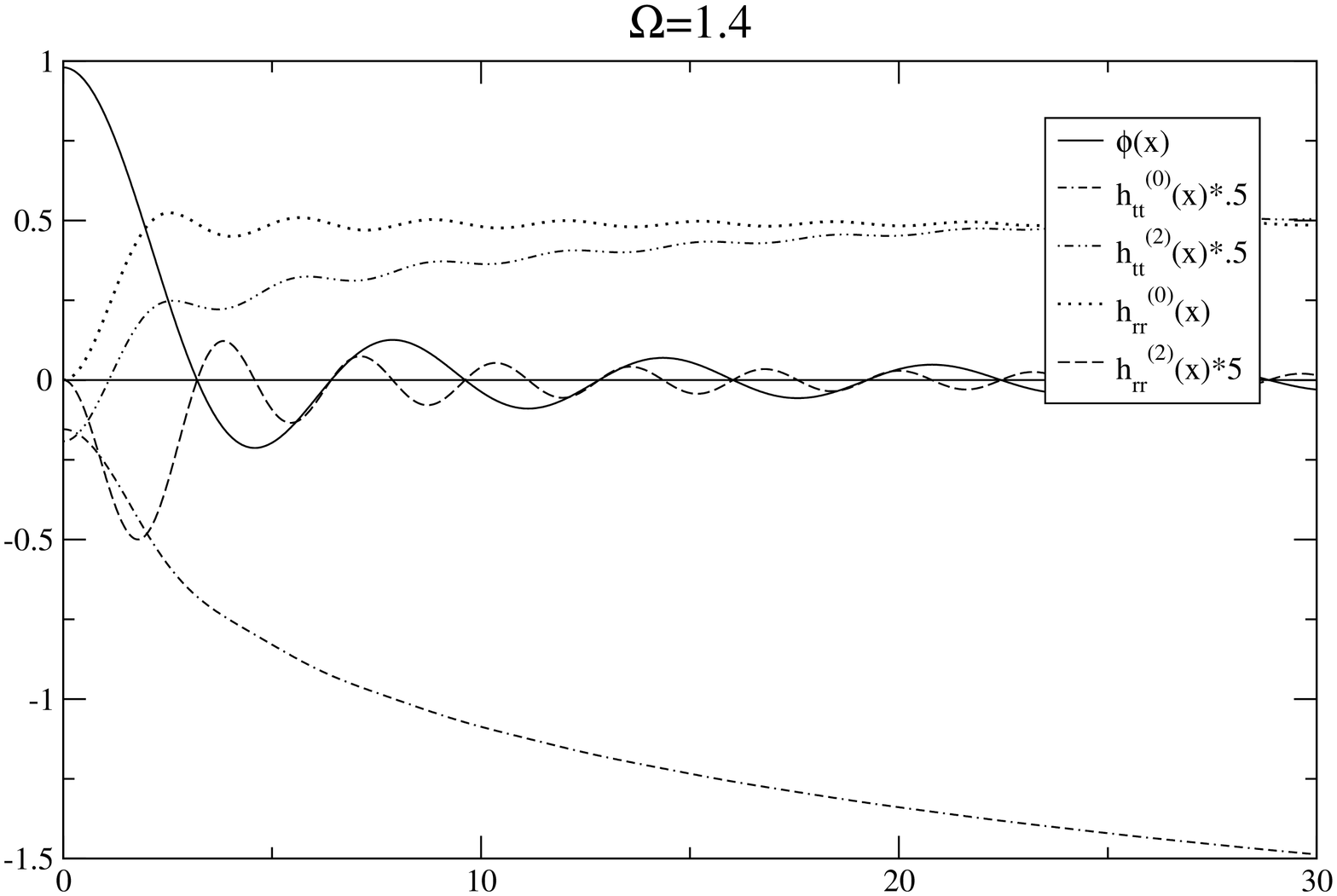}
\includegraphics[width=6.5cm, angle=270]{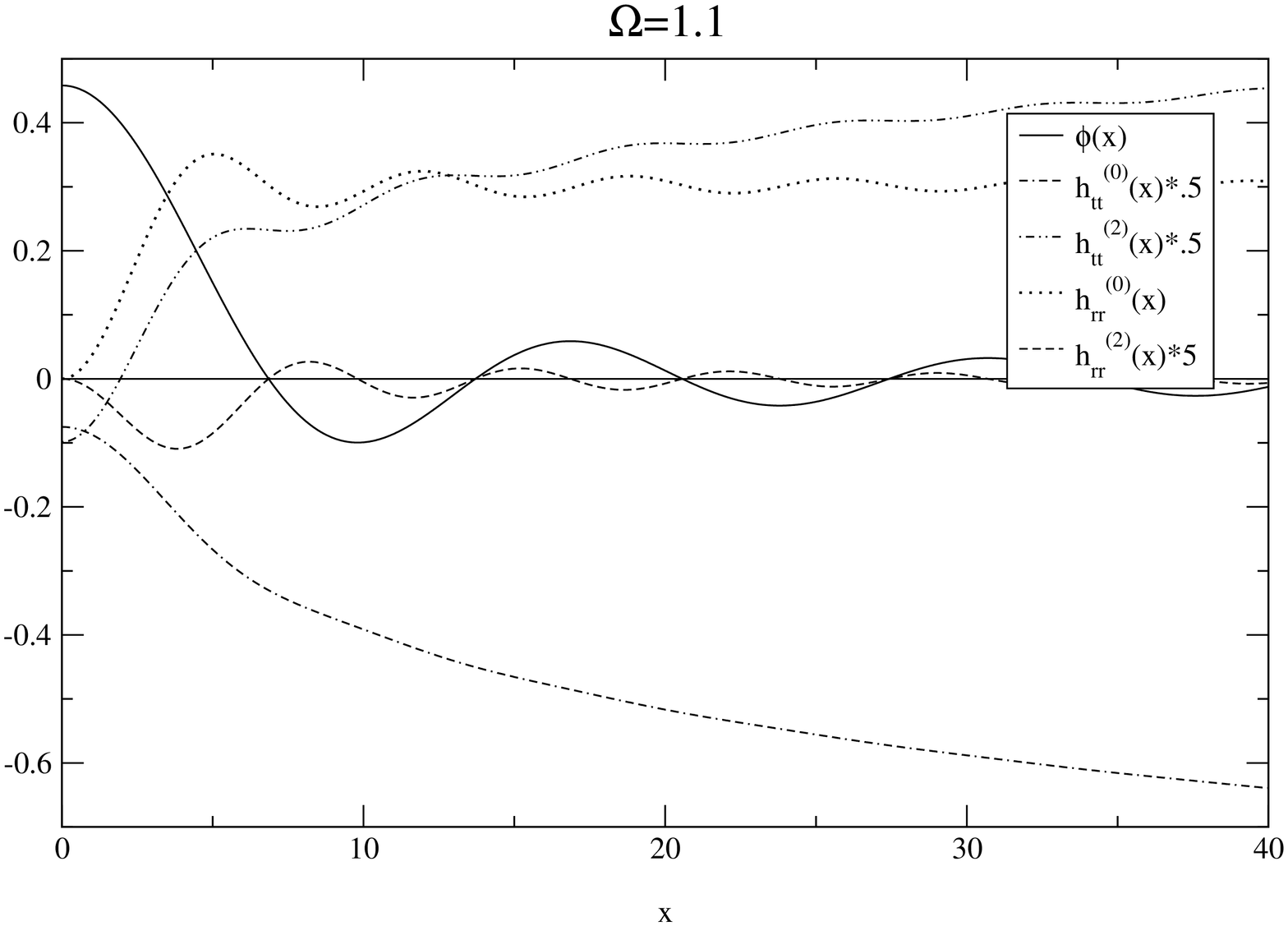}
\caption{\label{fig:solsr}Profiles of $\phi(x)$,
$h_{rr}^{(2)}(x)$, $h_{rr}^{(0)}(x)$,$h_{tt}^{(2)}(x)$ and
$h_{tt}^{(0)}(x)$ with $\phi_0=1$; see text for details. }
\end{center}
\end{figure}

Unique solutions for the EKG system will be obtained fixing the
$\phi_0$ and $\Omega$ parameters and the constants $C_{V02}$ and
$C_{V01}$ within the validity range of the approximation. As it is
known the potentials measurement does not have physical sense by
themselves, then unique solutions will be determined through
metric dependent observable quantities. Using the expressions
given by Eqs.~(\ref{eqs:hs}), we can obtain the perturbed metric
functions in terms of these solutions. In Fig. \ref{fig:solsr} we
present a plot of these metric perturbation functions, as well as
of the scalar field, for two values of $\Omega$.
\begin{figure}[htp]
\begin{center}
\includegraphics[width=6.5cm, angle=270]{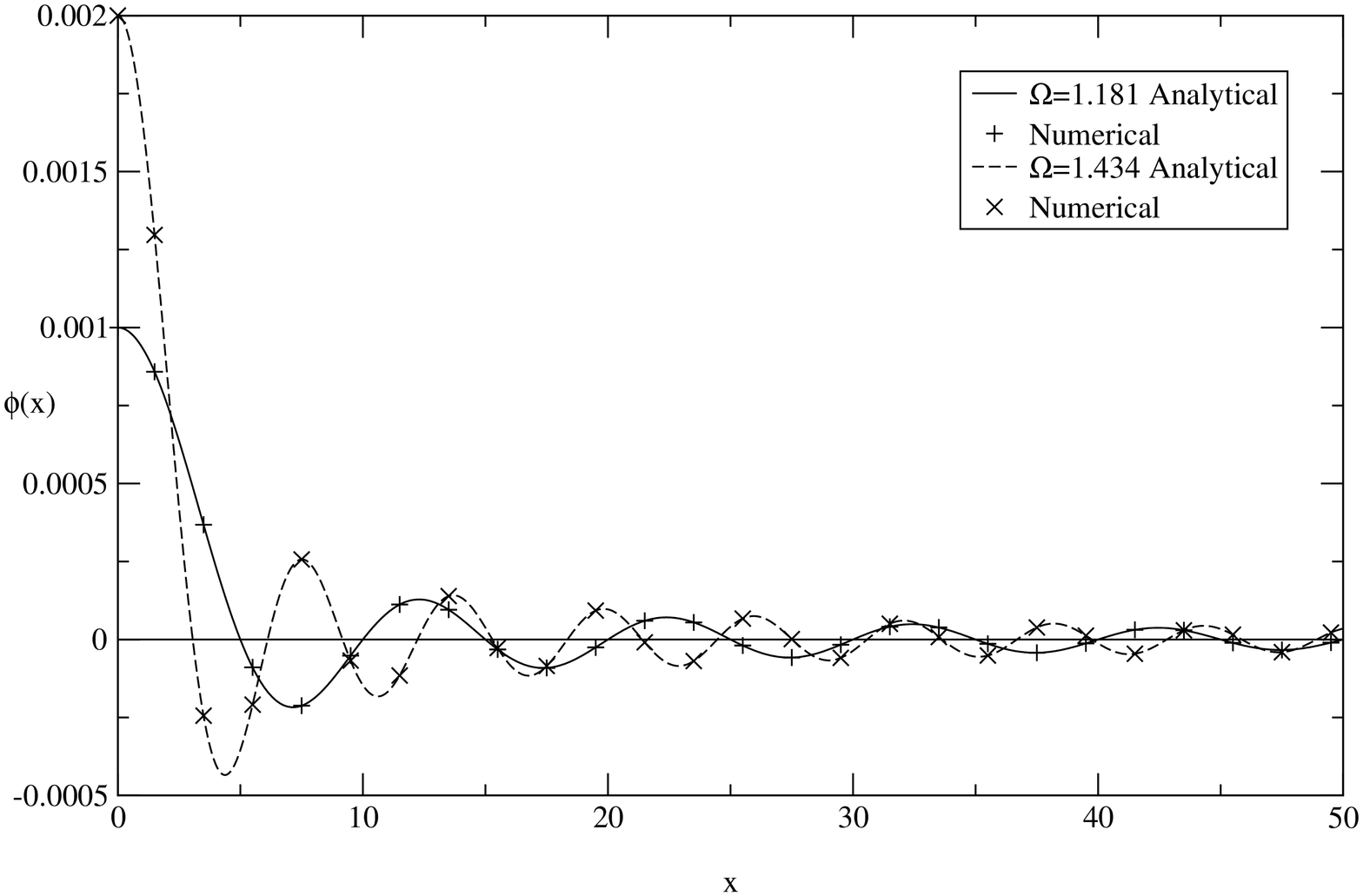}
\includegraphics[width=6.5cm, angle=270]{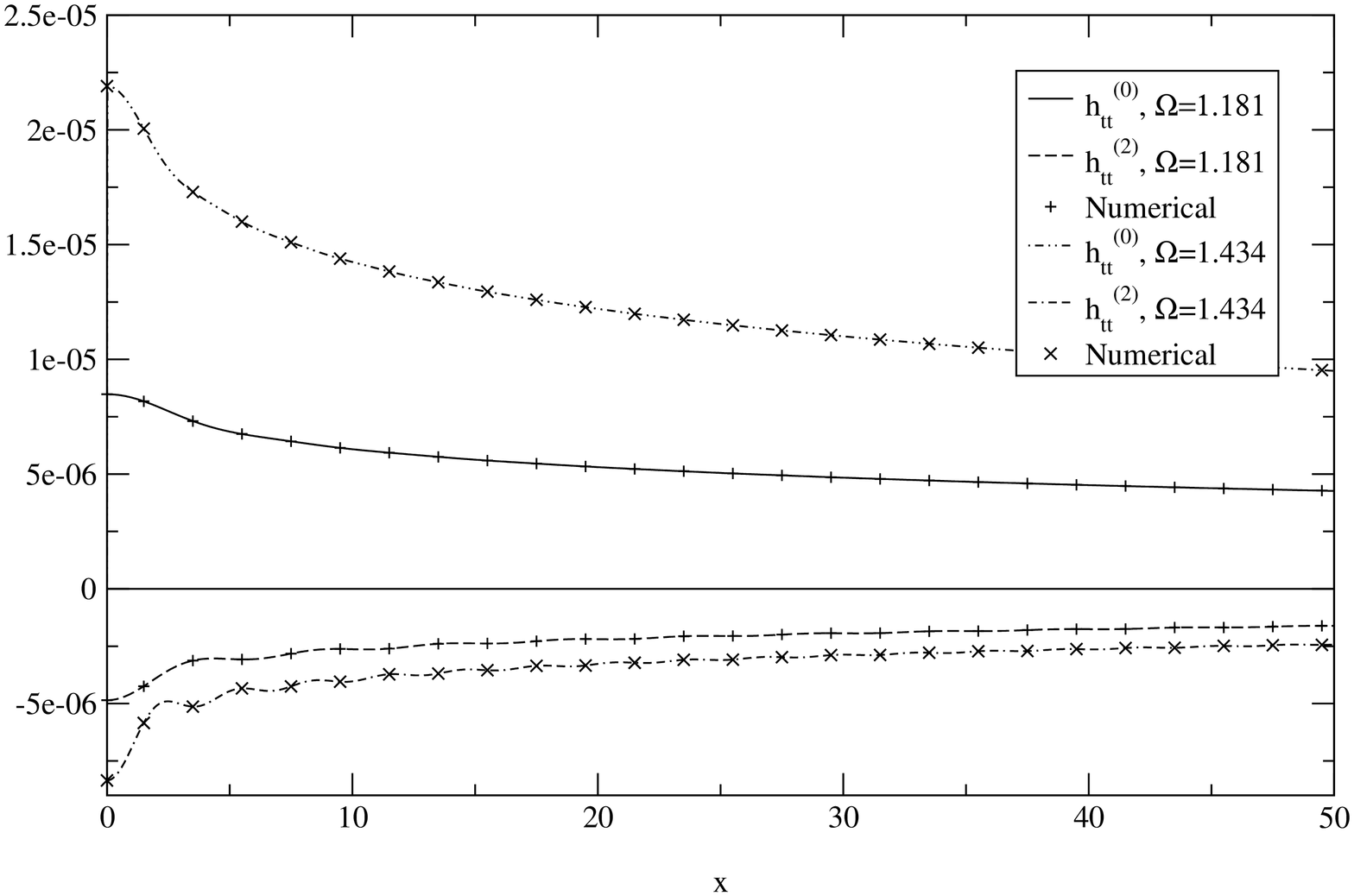}
\caption{\label{fig:phihtt} (Left) Spatial function $\phi$ of the
scalar field $\Phi$. Solid and dash lines are  analytical
solutions with $\phi_0=0.001/\sqrt{(\Omega^2-1)}= 0.001591$, and
$\phi_0=0.002/\sqrt{(\Omega^2-1)}=0.001946$, respectively. The
crosses are the corresponding numerical solutions with
$\phi_1(0)=0.001$, $\Omega=1.181008$ and $\phi_1(0)=0.002$,
$\Omega=1.433822$; see text for details. (Right) Spatial functions
$h_{tt}^{(0)}$ and $h_{tt}^{(2)}$ of the metric perturbations. }
\end{center}
\end{figure}
It is important to mention that the width of the validity region,
where (\ref{Weakfieldc}) is fulfilled, depends on $\Omega $ and
$\phi_0$. This is because it is the factor $\phi_0^2$ which
modulate the perturbations, (see (\ref{sols})). What it is not
evident until the solutions (\ref{Einsteins}) are observed is that
the $\Omega$ value, independently from $\phi_0$, could make the
validity range width bigger. This is because in the logarithm
argument there is the expression $\sqrt{\Omega^2-1}$, then as
$\Omega$ is closer to one the logarithm terms rise more slowly.

The order of magnitude for the other parameter $\phi_0$ in the
metric perturbations, can be naturally determined from the
asymptotic value taken by $h_{rr}$, which is reached very close to
the origin
\begin{equation}
\lim_{x\to \infty}|h_{rr}|=\frac{\phi^2_0}{4}\Omega^2.
\end{equation}
As $\Omega$ is nearly $1$ the $h_{rr}$ magnitude is given by
$\phi_0$. It is well known that for week field systems like our
Solar System the metric perturbations go like
$h_{\alpha\beta}\sim10^{-6}$. This value restricts our $\phi_0$ to
be $\phi_0\lesssim 10^{-3}$.

\subsection{Analytical Solutions vs Numerical Solutions}

Analytic solutions $\phi_0$, $h_{rr}^{(2)}(x)$, $h_{rr}^{(0)}(x)$,
$h_{tt}^{(2)}(x)$ and $h_{tt}^{(0)}(x)$ are shown in
Fig.~\ref{fig:solsr}. The value of $\phi_0$ is $1$ in both plots.
As already was noticed, the value of $\Omega$ characterizes each
family of solutions. Mainly $\Omega$ determines the wave length of
$\phi$ and the increase rate of $h_{tt}^{(0)}$ and $h_{tt}^{(2)}$;
as $\Omega$ is closer to $1$ this rate is smaller. These
characteristics are shown in Fig.~\ref{fig:solsr}.
\begin{figure}[htp]
\begin{center}
\includegraphics[width=6.5cm, angle=270]{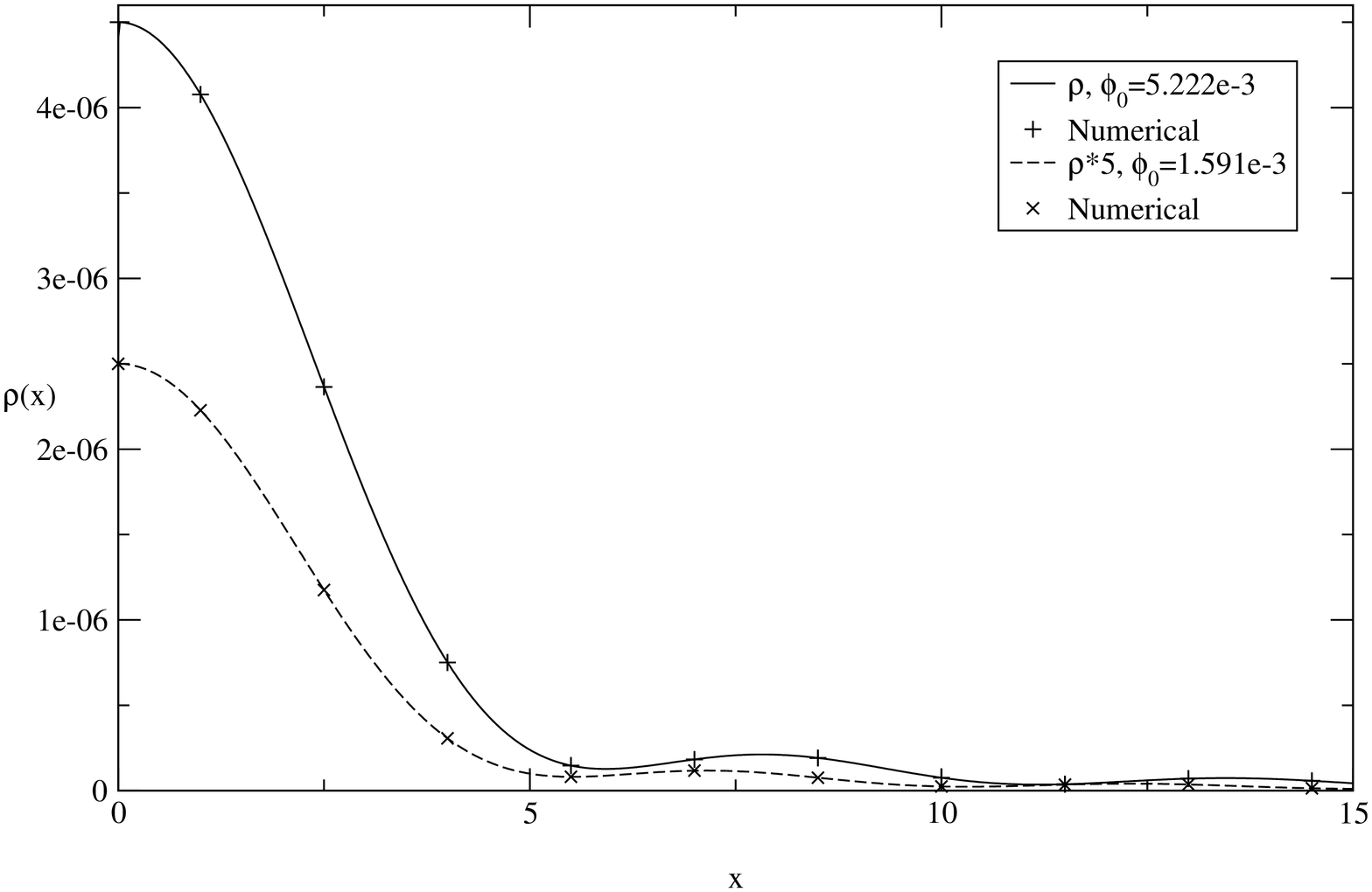}
\includegraphics[width=6.5cm, angle=270]{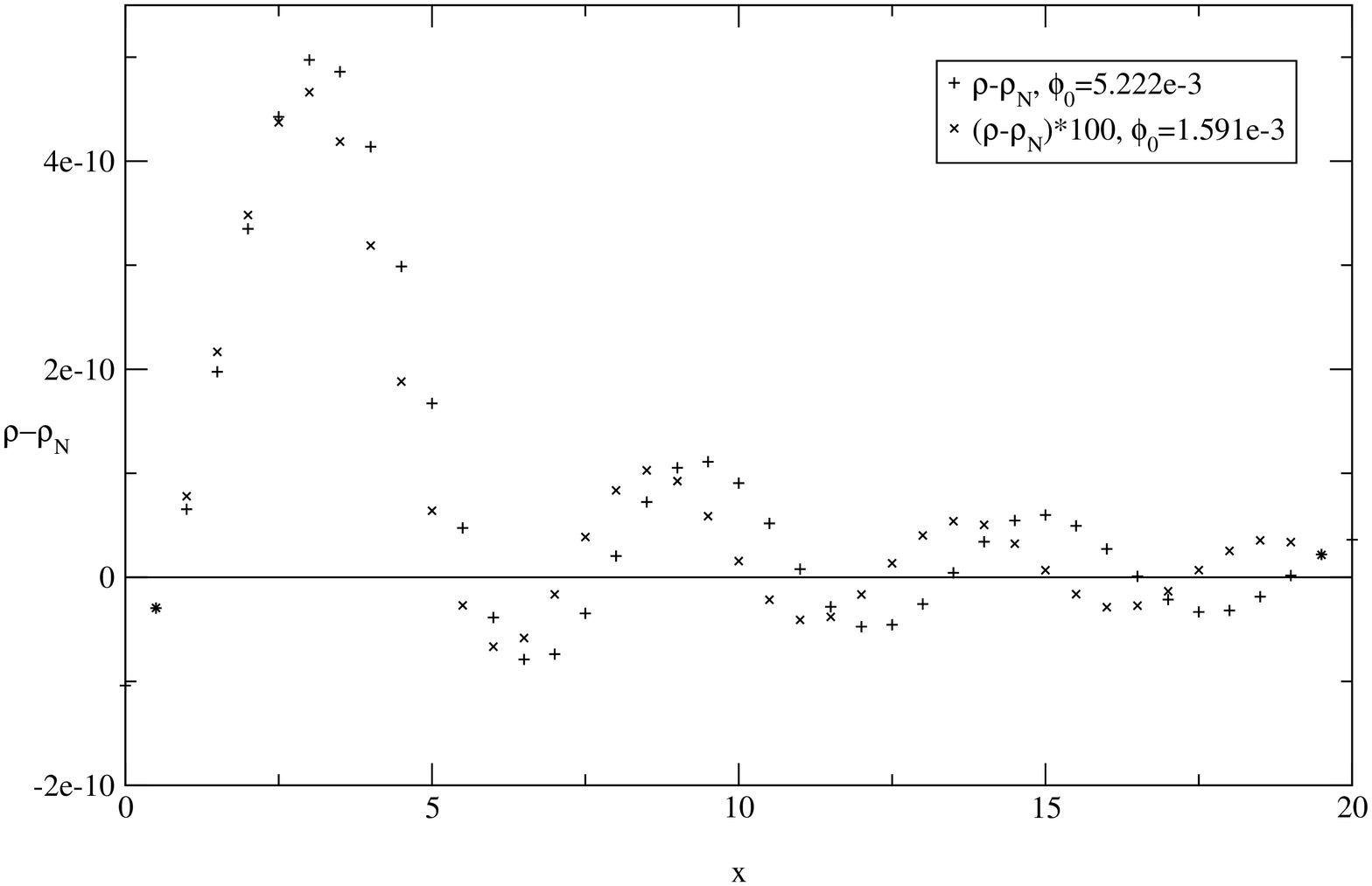}
\caption{\label{fig:rho} (Left) Energy-Momentum density profiles
for two scalar field configurations one with
$\phi_0=1.1591\,\,10^{-3}$ and $\Omega=1.181008$ and the other
with $\phi_0=5.2225\,\,10^{-3}$ and $\Omega=1.153246$. (Right)
Difference between density profiles from the exact EKG equations
($\rho_N$) and the density computed in the flat space-time
($\rho$); see text for details. }
\end{center}
\end{figure}

The exact EKG equations in spherical symmetry with a quadratic
potential, were solved numerically in Ure{\~n}a-L{\'o}pez 2002,
and
 Ure{\~n}a-L{\'o}pez et al 2002)
and found the so called oscillatons. In those works boundary
conditions are determined by requiring non-singular and
asymptotically flat solutions, for which the EKG become an
eigenvalue problem. The free  eigenvalue is the scalar field's
central value $\phi_1(x=0)$ which labels the particular
equilibrium configuration, and the fundamental frequency $\Omega$
is an output value. In those works it is also noted that weak
gravity fields are produced by oscillatons with $\phi_1(x=0)\ll
1$. In Fig.~\ref{fig:phihtt} we compare some of these numerical
solutions (NS) with the analytical solutions (AS) within a central
region. The constant values of the AS are fixed to fit better the
NS inside the weak field validity range. From these plots we can
conclude that our solutions are a very good approximation for the
exact EKG equations in the weak field limit. The principal
advantage of this approximation is the analytical description of
the solutions.

\section{Scalar Field as Dark Matter: Halo Density Profile}
\label{sect:sf}

In this section we explore whether or not the scalar field
could account as the galactic DM halos. Specifically we
compare the SFDM model density profile and the profiles inferred
throughout the rotation curves profiles of galaxies which are
mostly formed by DM.

As long as we are concerned with perturbations of the flat space-time
due to the scalar field, we do not consider the baryonic
matter gravitational effects, we expect that our approximation will be
better suited for galaxies with very small baryonic component.

We will compare the energy-momentum density for the scalar field
given by Eq.~(\ref{tensor1}), in the relativistic weak limit
approximation, where for the metric functions,
Eq.(\ref{perturbations}), we use the solutions to the
perturbations given by Eqs.~(\ref{eq:VU},\ref{Einsteins}).
\begin{figure}[htp]
\begin{center}
\includegraphics[width=3.0cm, angle=270]{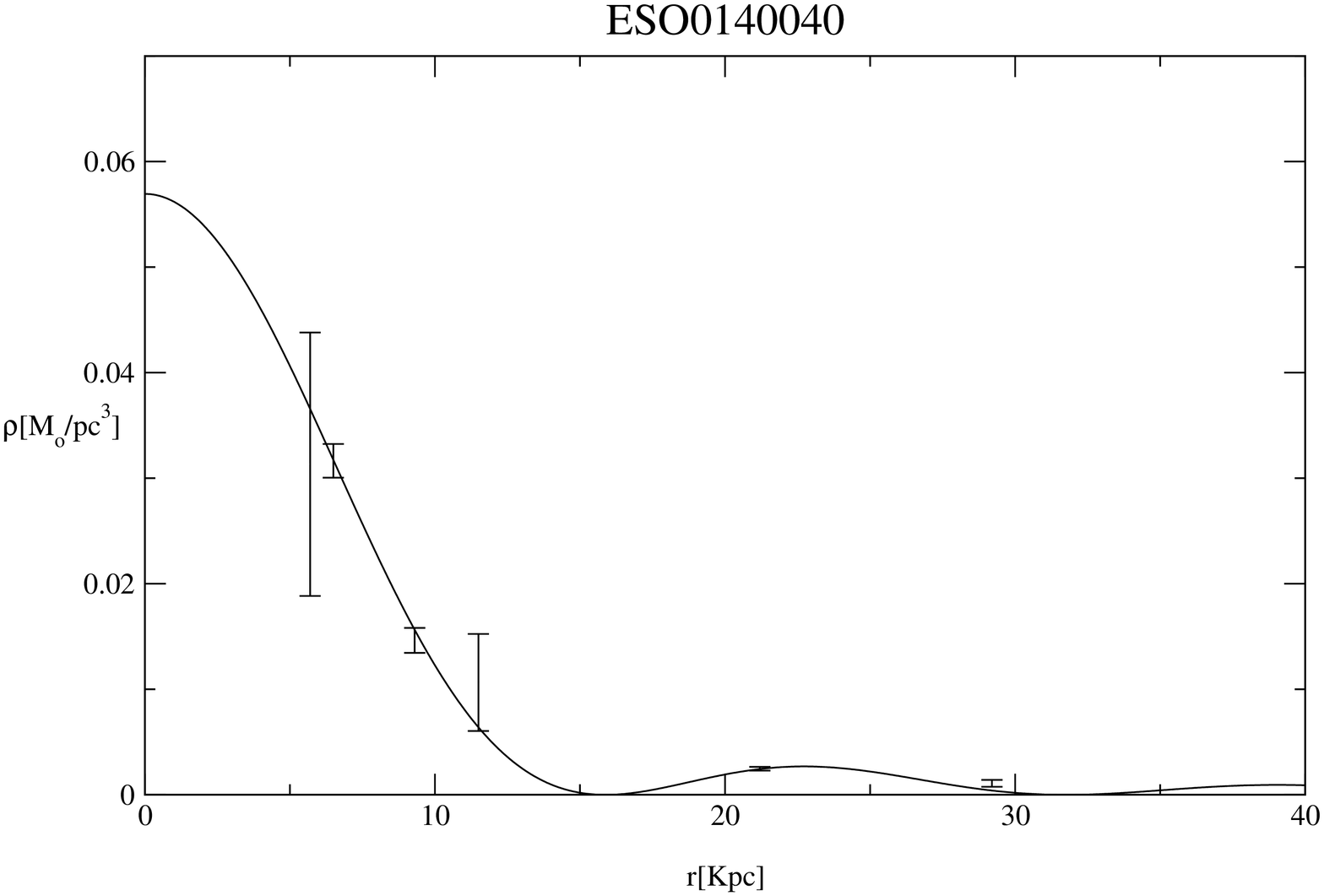}
\includegraphics[width=3.0cm, angle=270]{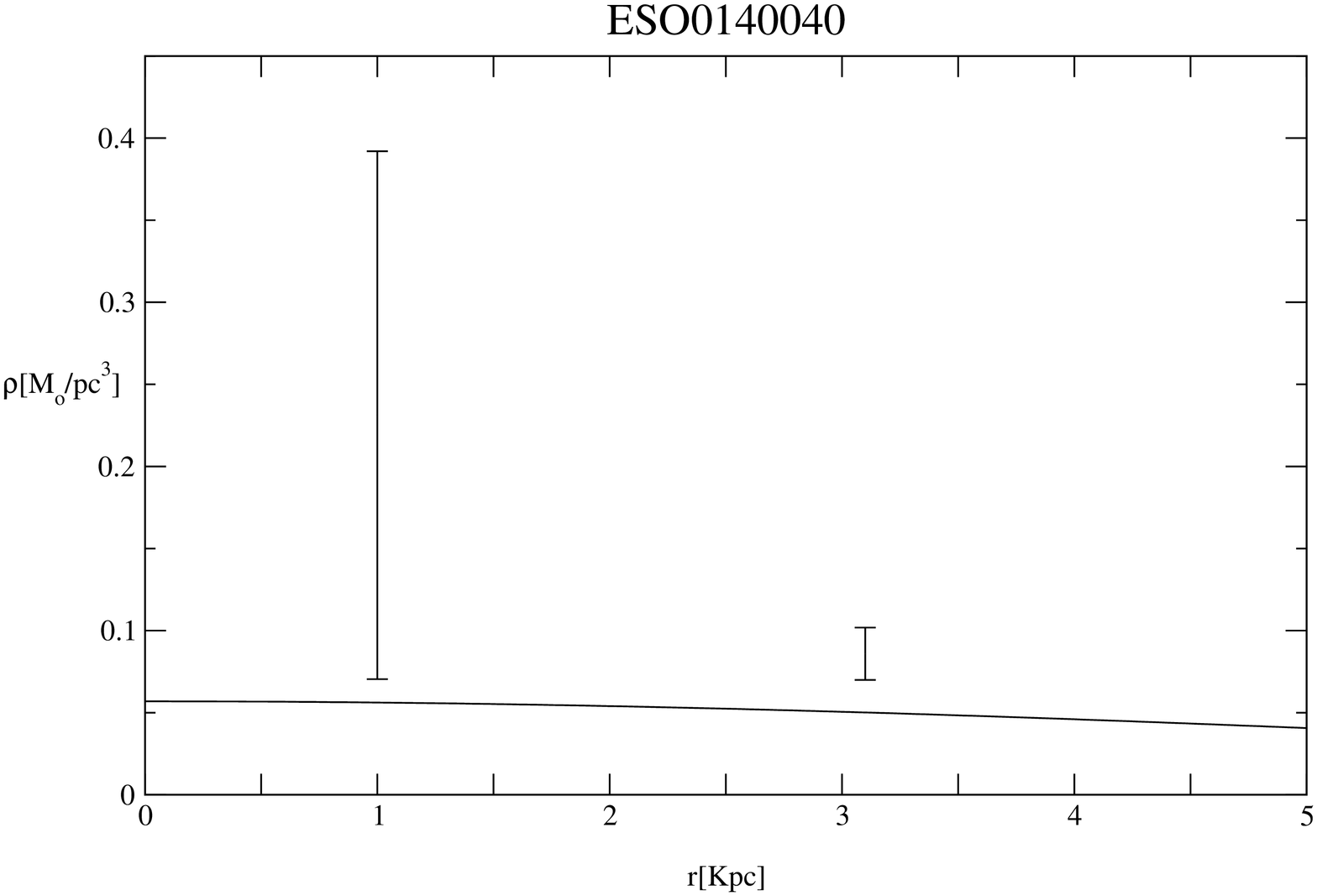}

\includegraphics[width=3.0cm, angle=270]{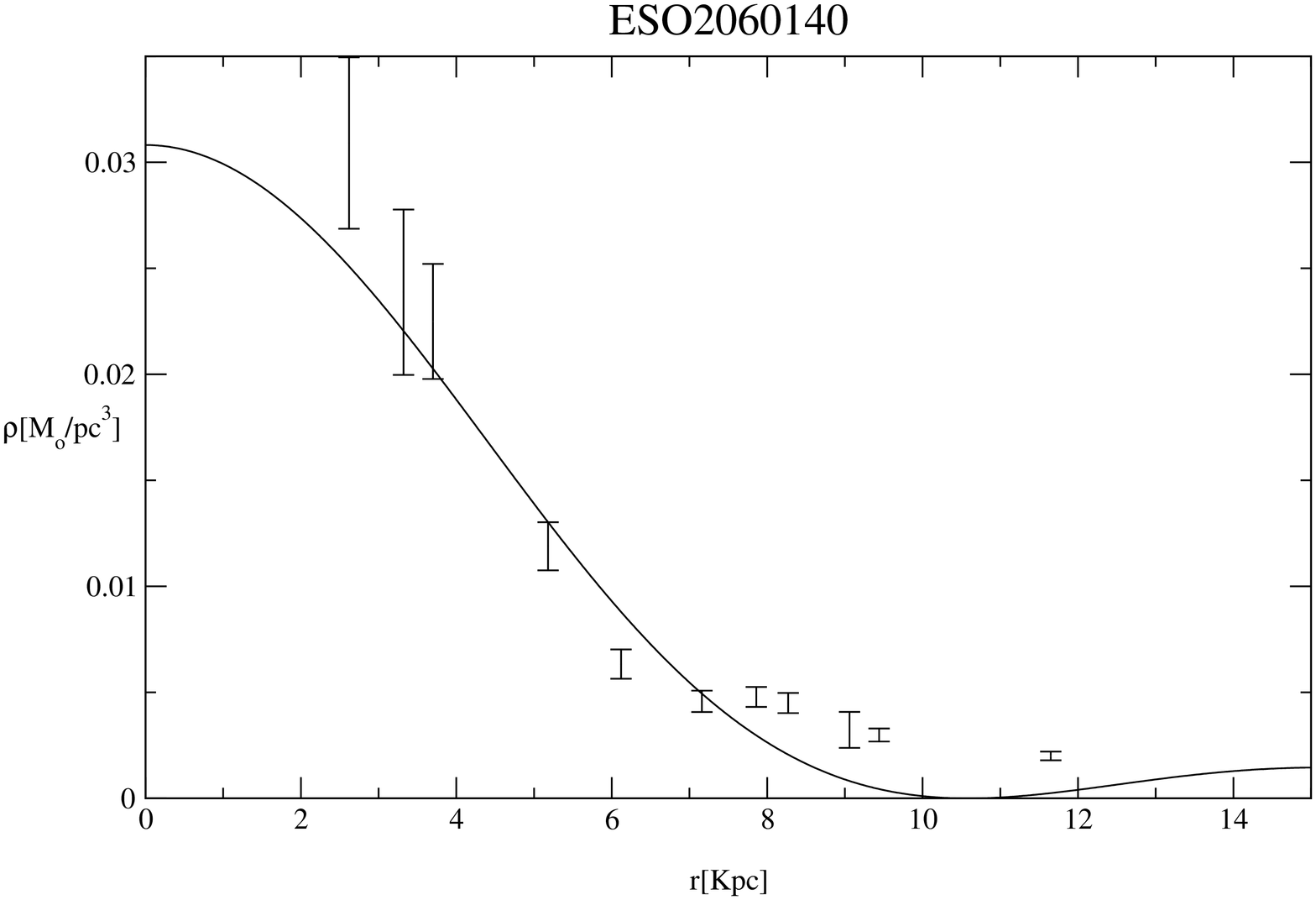}
\includegraphics[width=3.0cm, angle=270]{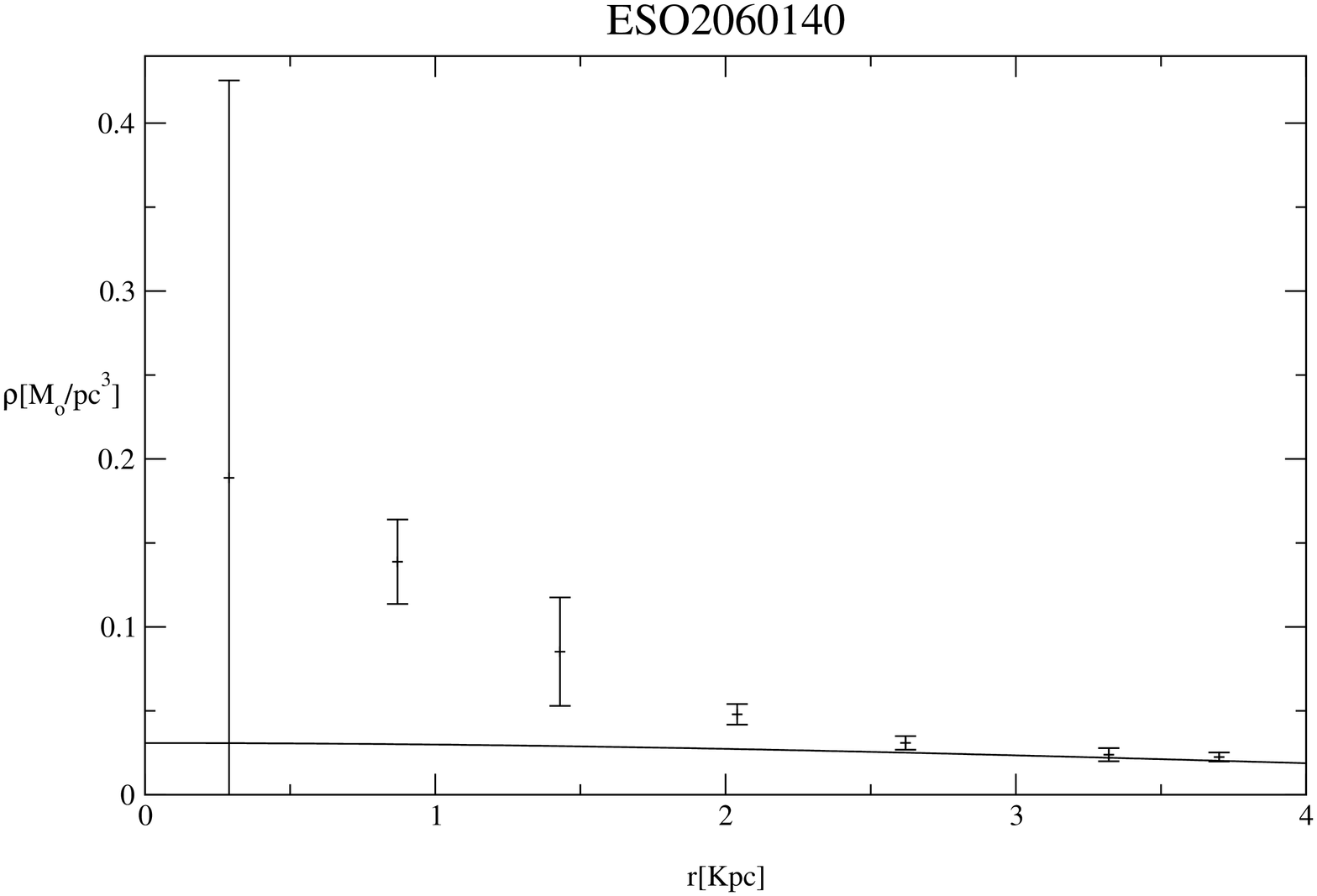}
\caption{\label{fig:G01} Density profile fits for the galaxies
ESO0140040, ESO2060140, the right panel is a zoom of its central
region, it is plotted in order to visualize complete error bars.
The horizontal line is in $kpc$, vertical line in $M_{\odot}/pc^3$.
See text and Table ~\ref{Table1} for fit details.}
\end{center}
\end{figure}
This is consistent with the fact that gravity does not modify the
scalar field behavior. This approximation in the weak
gravitational field limit is very good as we can see in Fig.
~\ref{fig:rho}. In those plots we show the energy-momentum density
profiles for two scalar field configurations with different
maximum amplitudes at the origin $\phi(0)=\phi_0\sqrt{\Omega^2-1}$.
It is important to notice that as $\phi(0)$ decreases the
gravitational field gets weaker, and the difference between the
density from the complete EKG equations and from our approximation
is smaller.
\begin{figure}[htp]
\begin{center}
\includegraphics[width=3.0cm, angle=270]{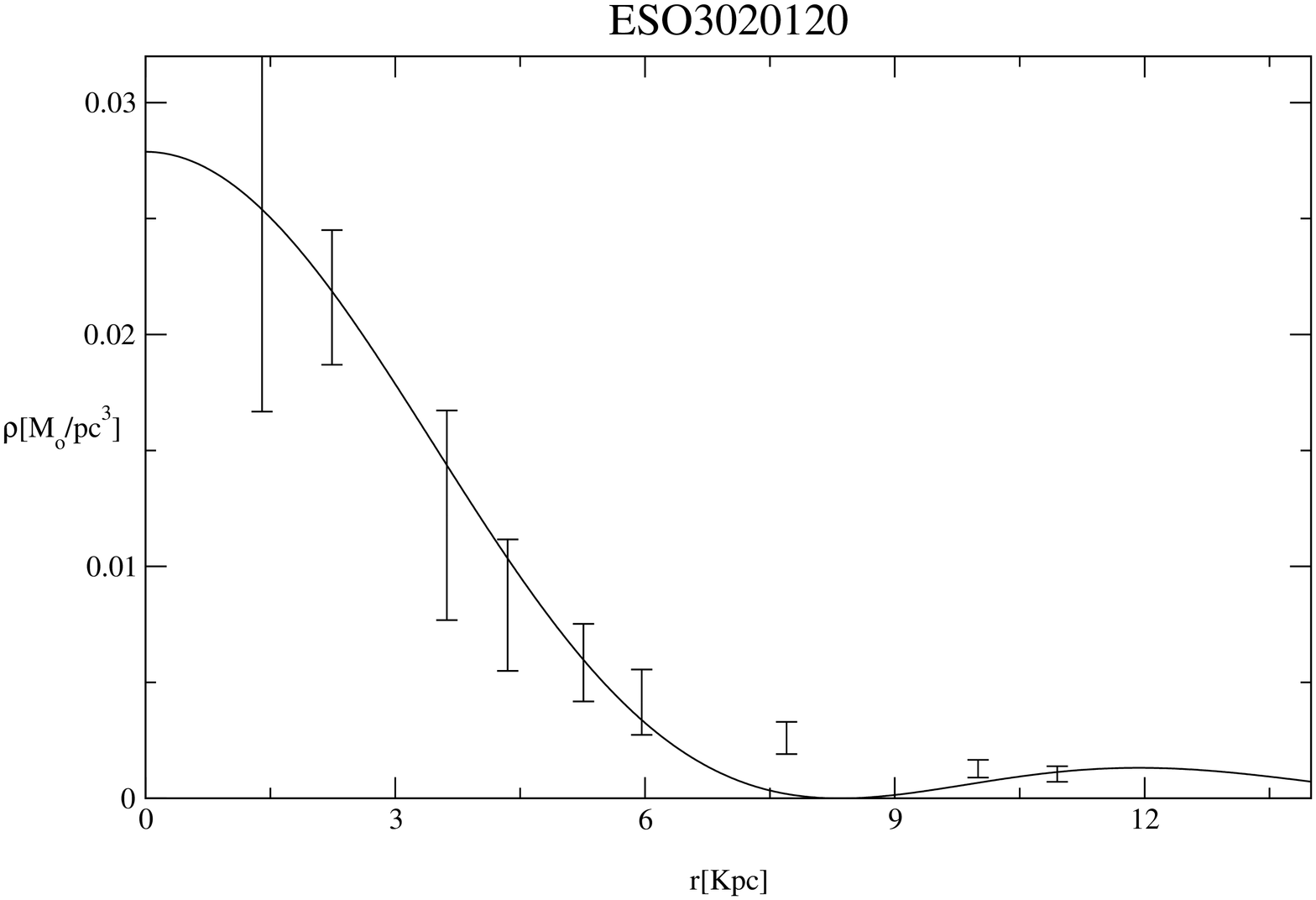}
\includegraphics[width=3.0cm, angle=270]{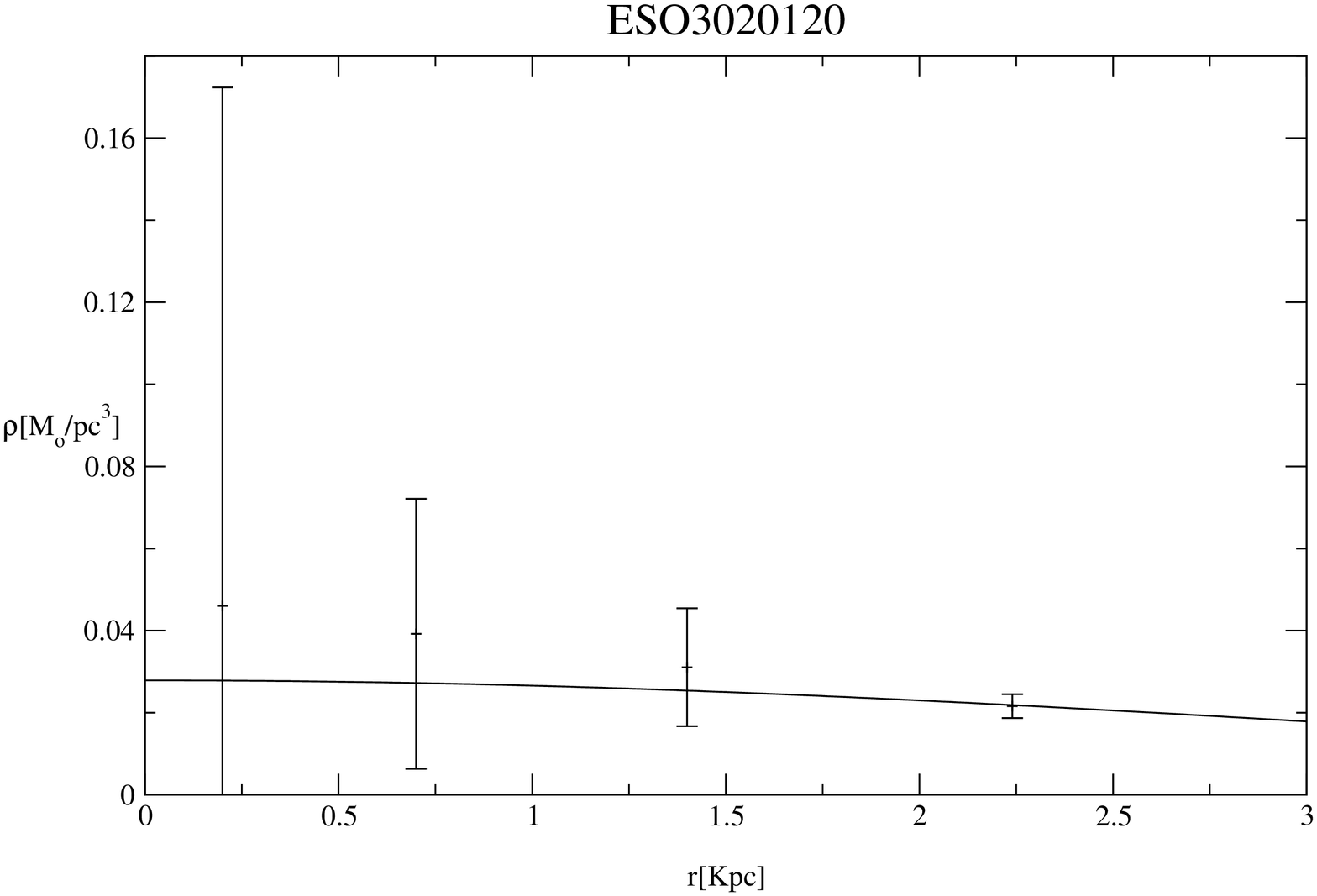}

\includegraphics[width=3.0cm, angle=270]{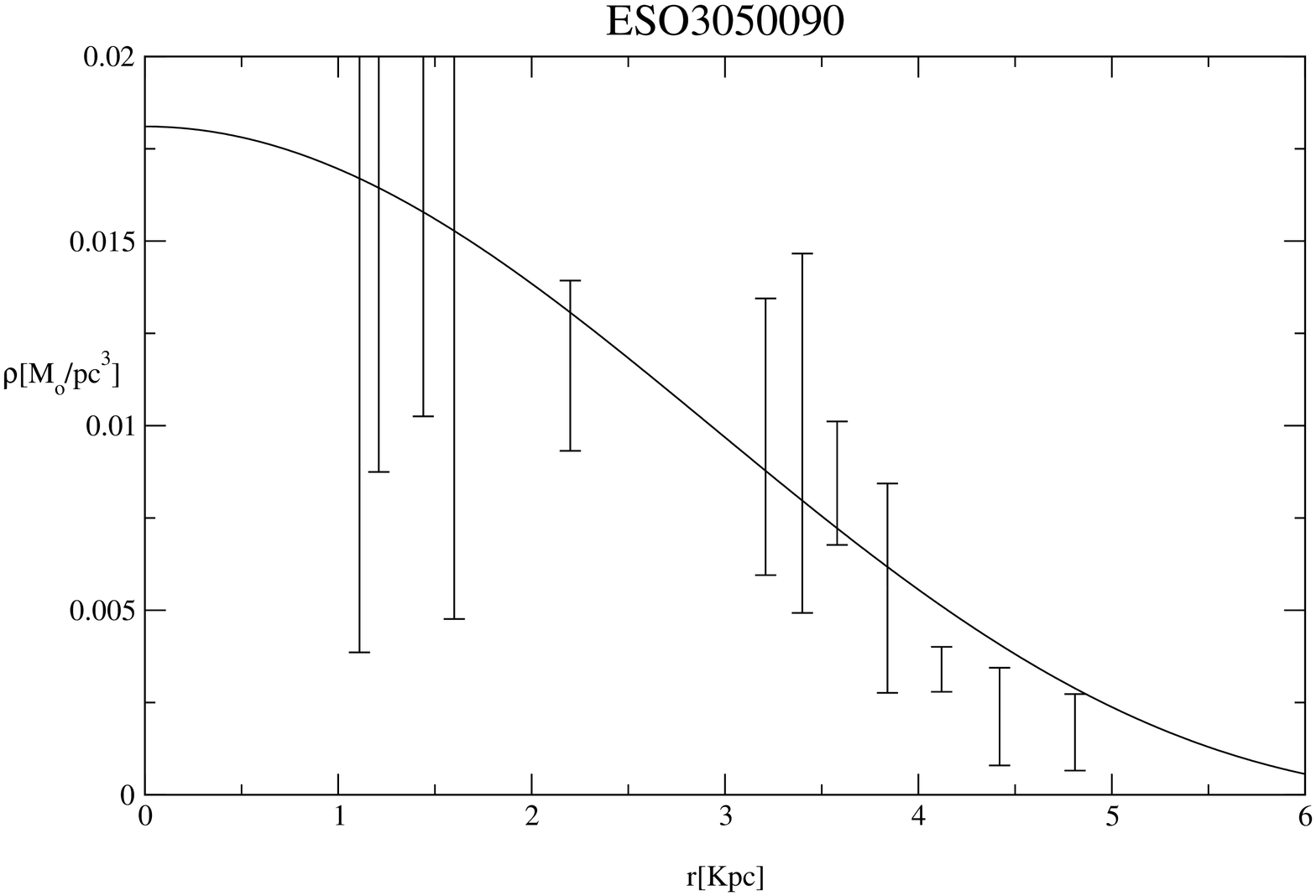}
\includegraphics[width=3.0cm, angle=270]{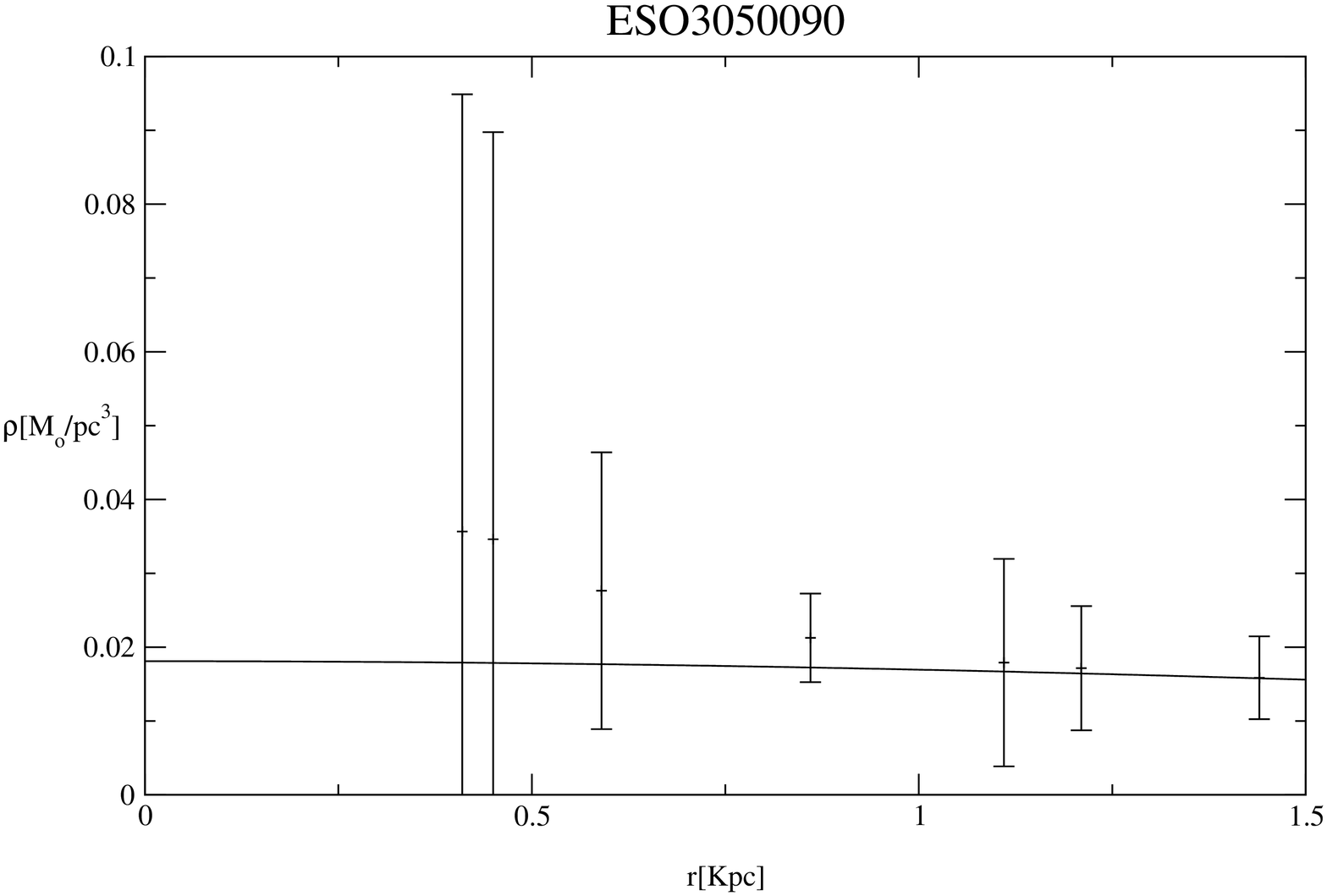}
\caption{\label{fig:G02}The same as Fig. \ref{fig:G01}, but
 for the galaxies ESO3020120, ESO3050090.}
\end{center}
\end{figure}

The density profile fits allow us to obtain an estimation of the
parameters at the galactic level: the fundamental frequency $\Omega$
and the scalar field constant $\phi_0$. The third parameter
involved in the density profiles is the scalar field mass, we will
fix it to be $m=10^{-23}$ eV. This value was fitted for the SFDM
model from cosmological observations in Matos \&
Ure{\~n}a-L{\'o}pez 2001.
\begin{figure}[htp]
\begin{center}
\includegraphics[width=3.0cm, angle=270]{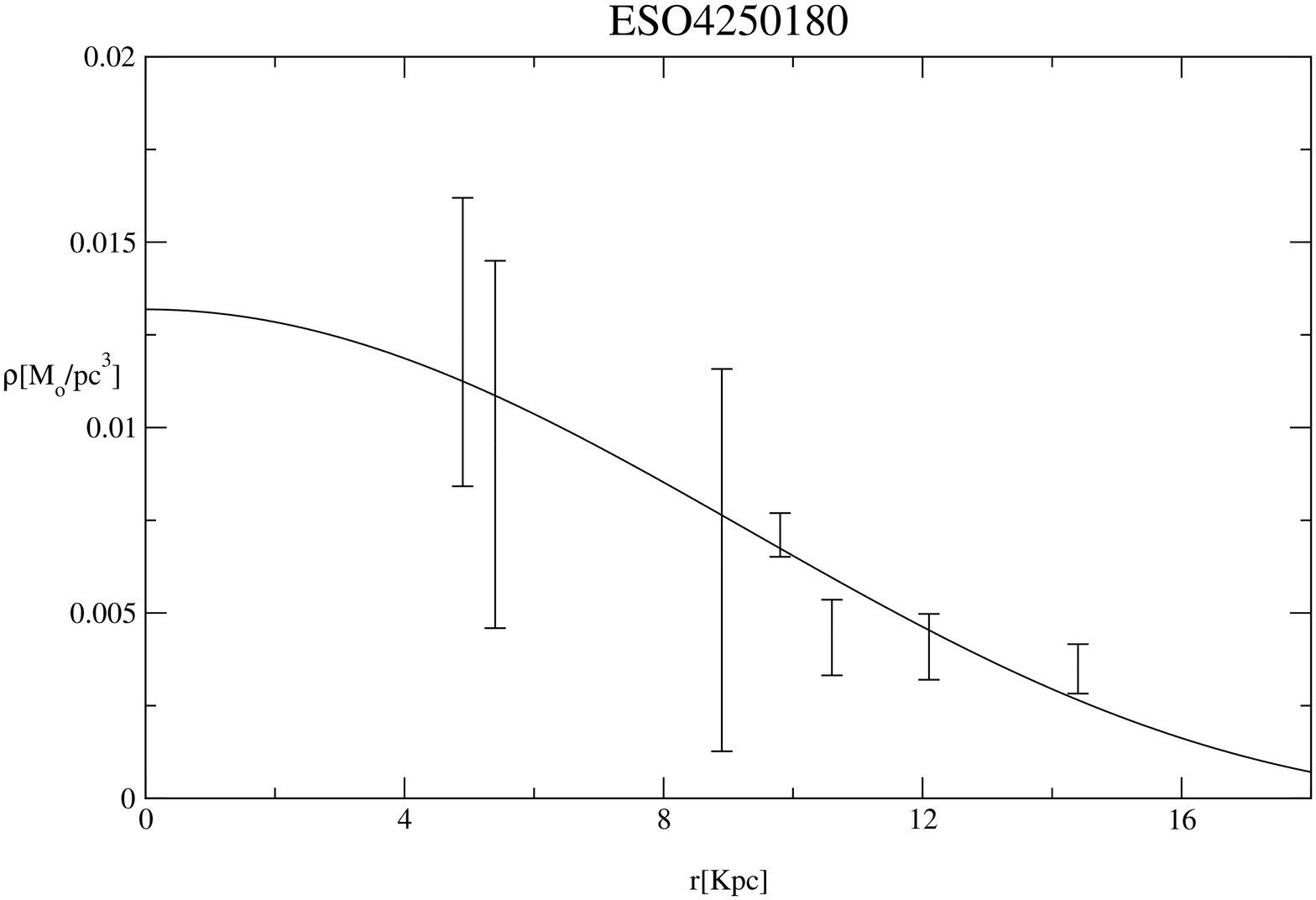}
\includegraphics[width=3.0cm, angle=270]{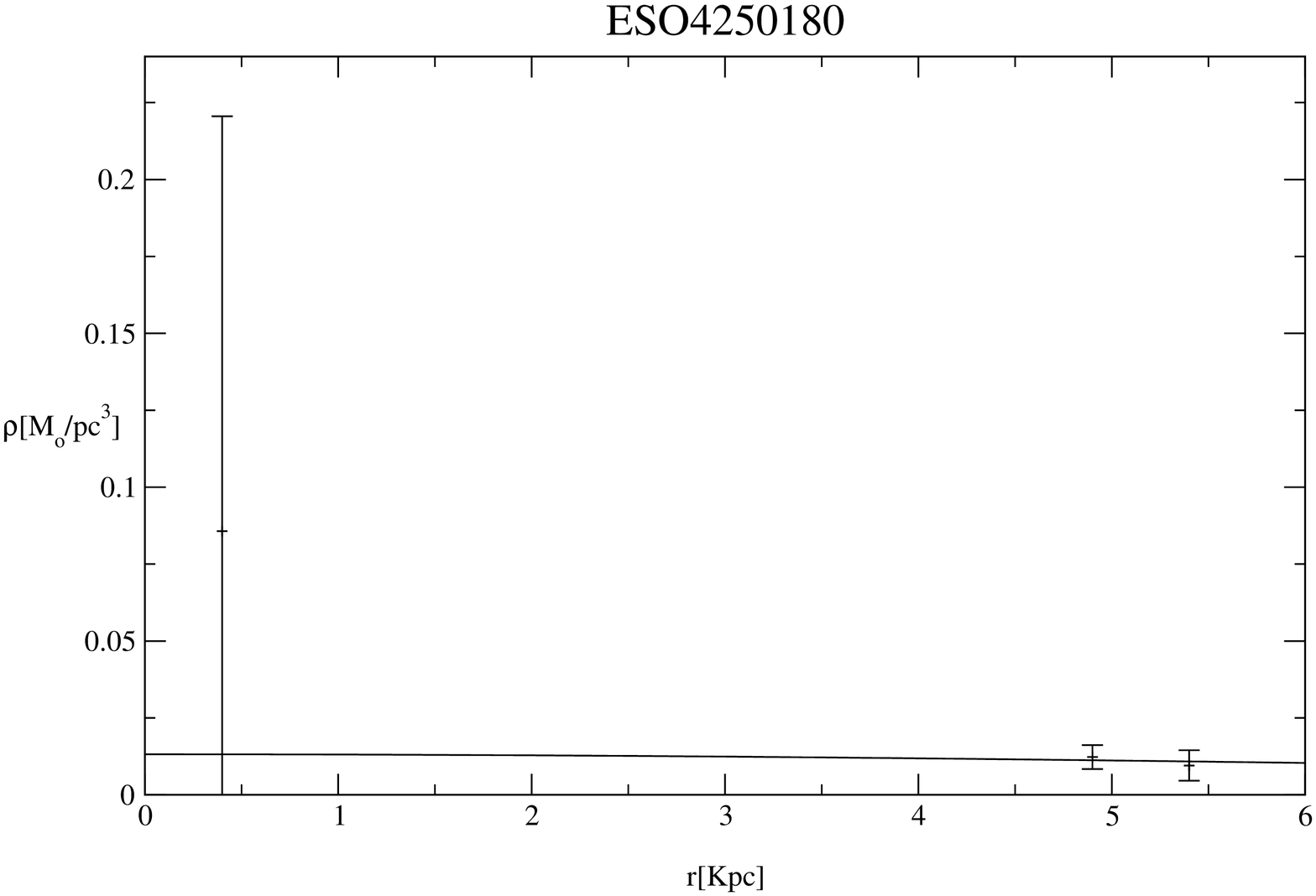}

\includegraphics[width=3.04cm, angle=270]{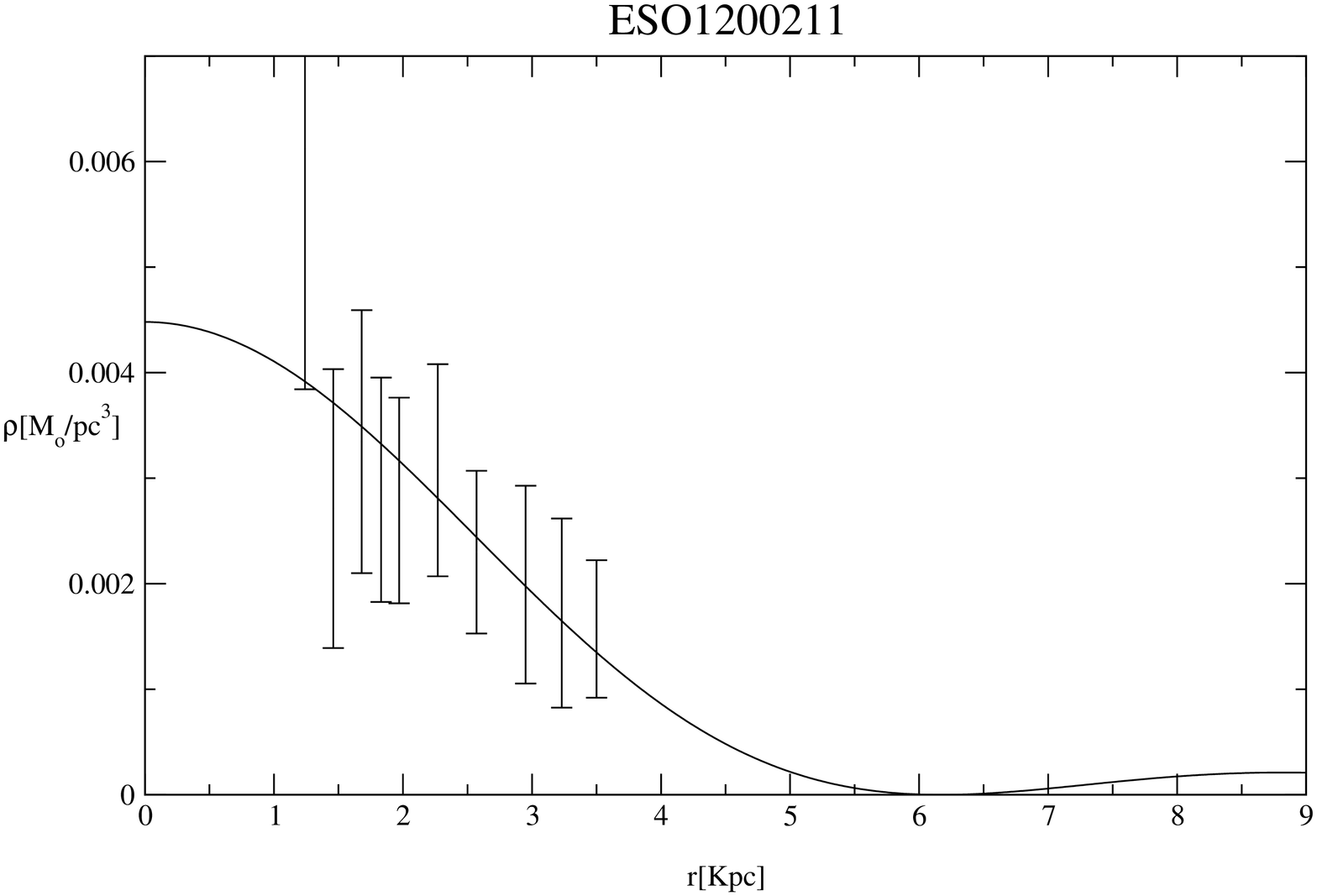}
\includegraphics[width=3.0cm, angle=270]{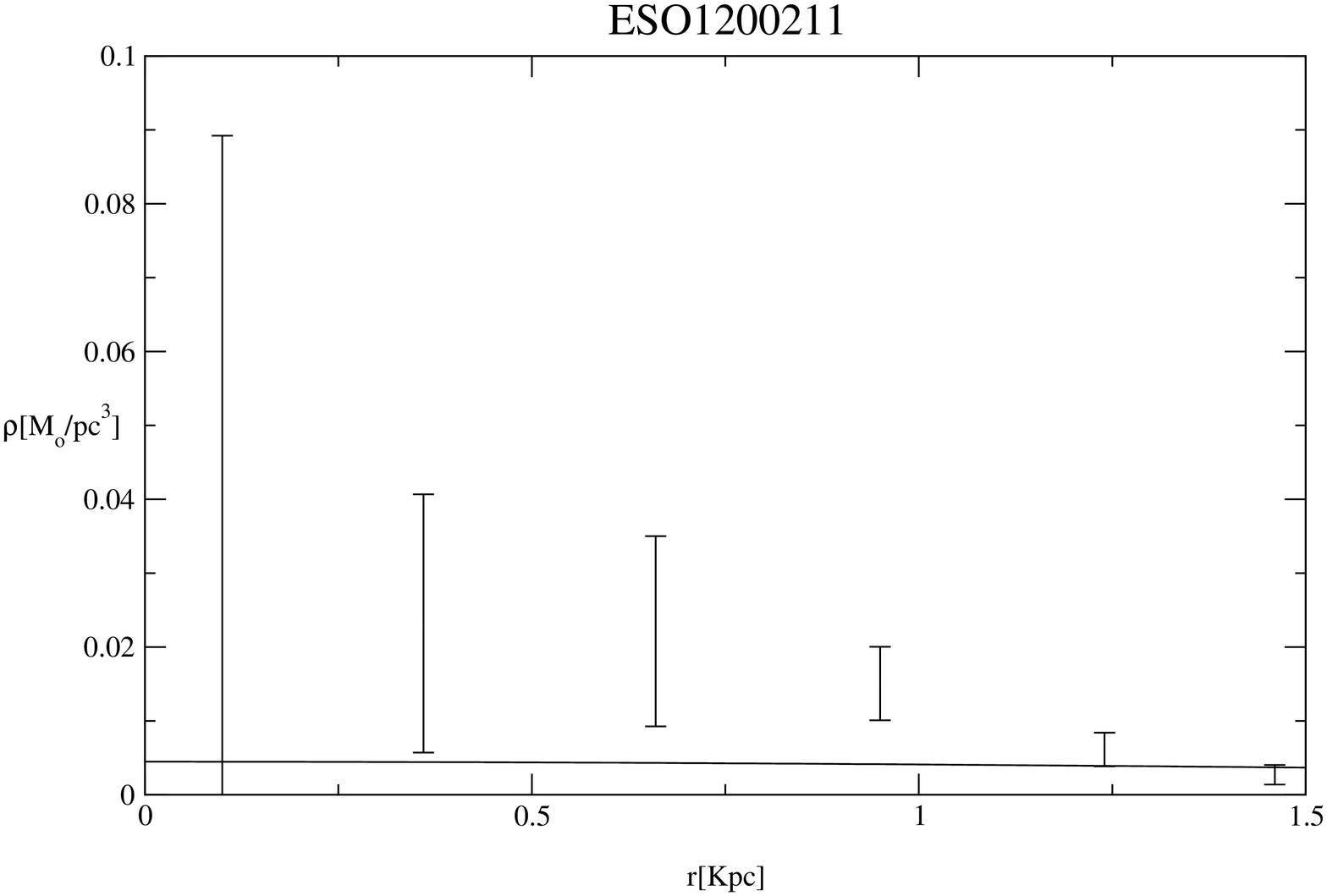}
\caption{\label{fig:G03}The same as Fig. \ref{fig:G01}, but
 for the galaxies ESO4250180, ESO1200211. }
\end{center}
\end{figure}
\subsection{Density Profile fits}

The first qualitative feature of the energy-density profile that
we want to emphasize is that in the central region it is non cuspy
(see Fig. \ref{fig:rho}). It is important to take into account
that instead of density profiles, rotation curves are the direct
observable for galaxies. Nevertheless, for galaxies dominated by
DM, their rotation curves could model the DM density profile more
trustfully. We choose a subset of galaxies from the set presented
in McGaugh 2001, the common characteristic for the selected
galaxies is that the luminous matter velocity contribution to the
rotation curves is almost null.

\begin{figure}[htp]
\begin{center}
\includegraphics[width=4.4cm, angle=270]{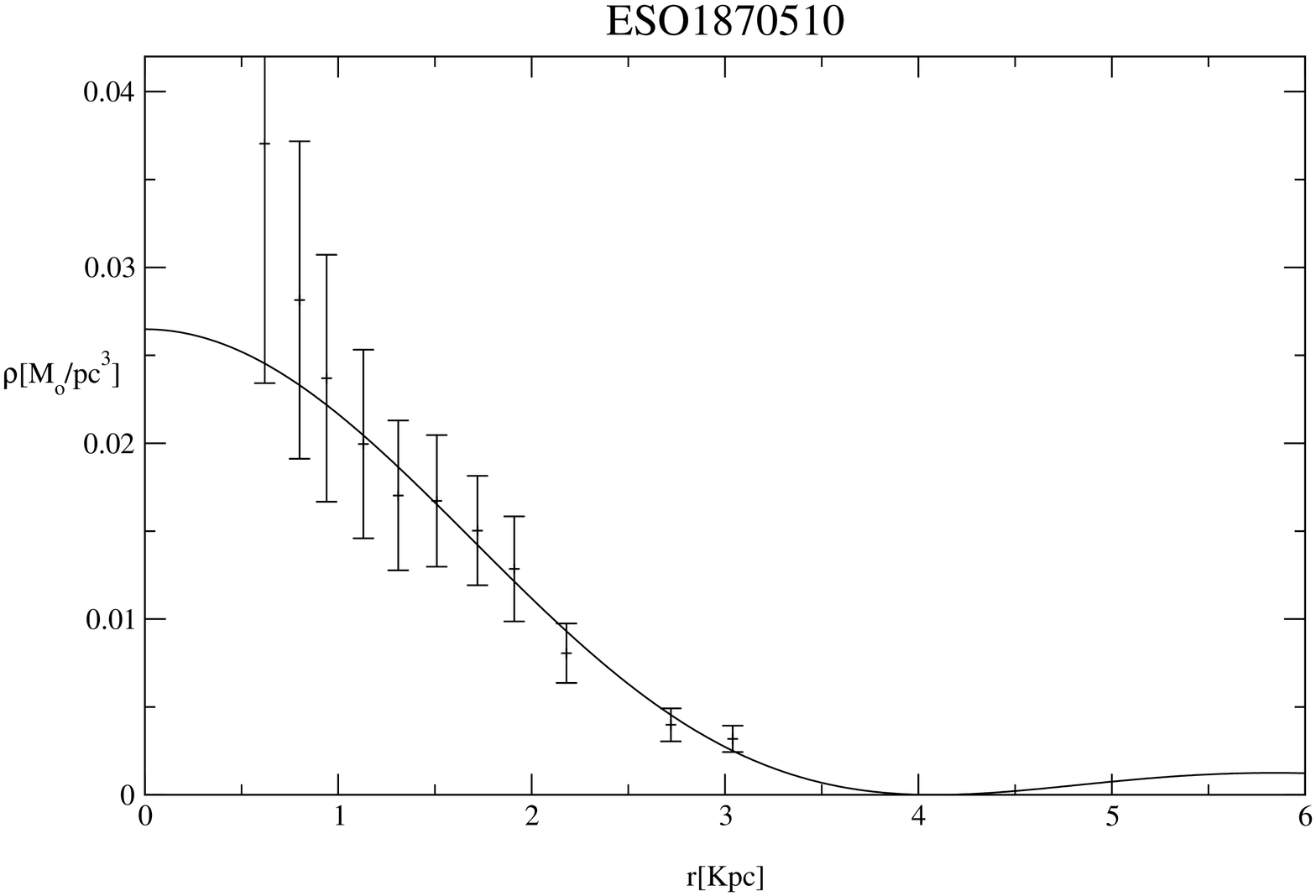}
\includegraphics[width=4.4cm, angle=270]{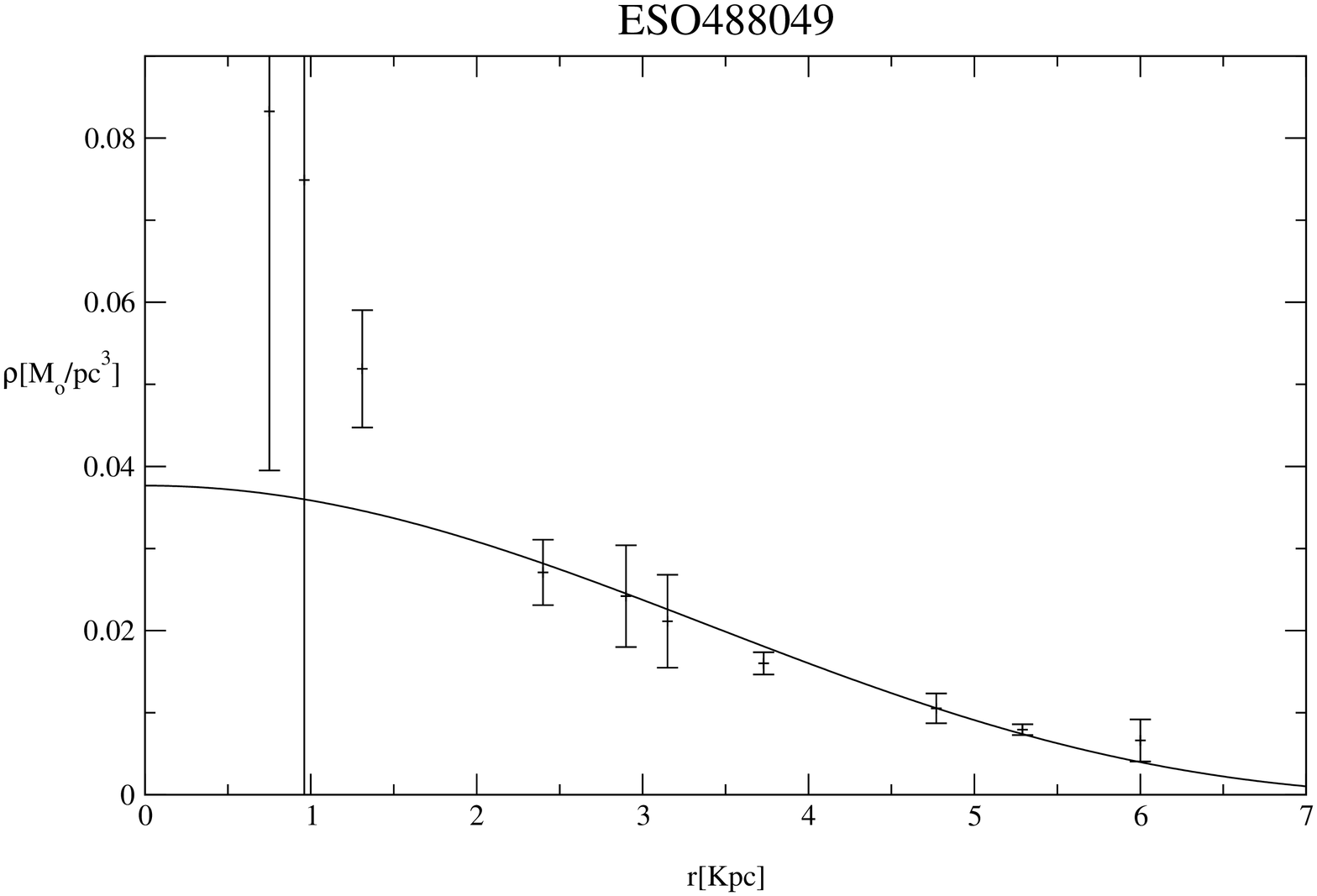}
\includegraphics[width=4.4cm, angle=270]{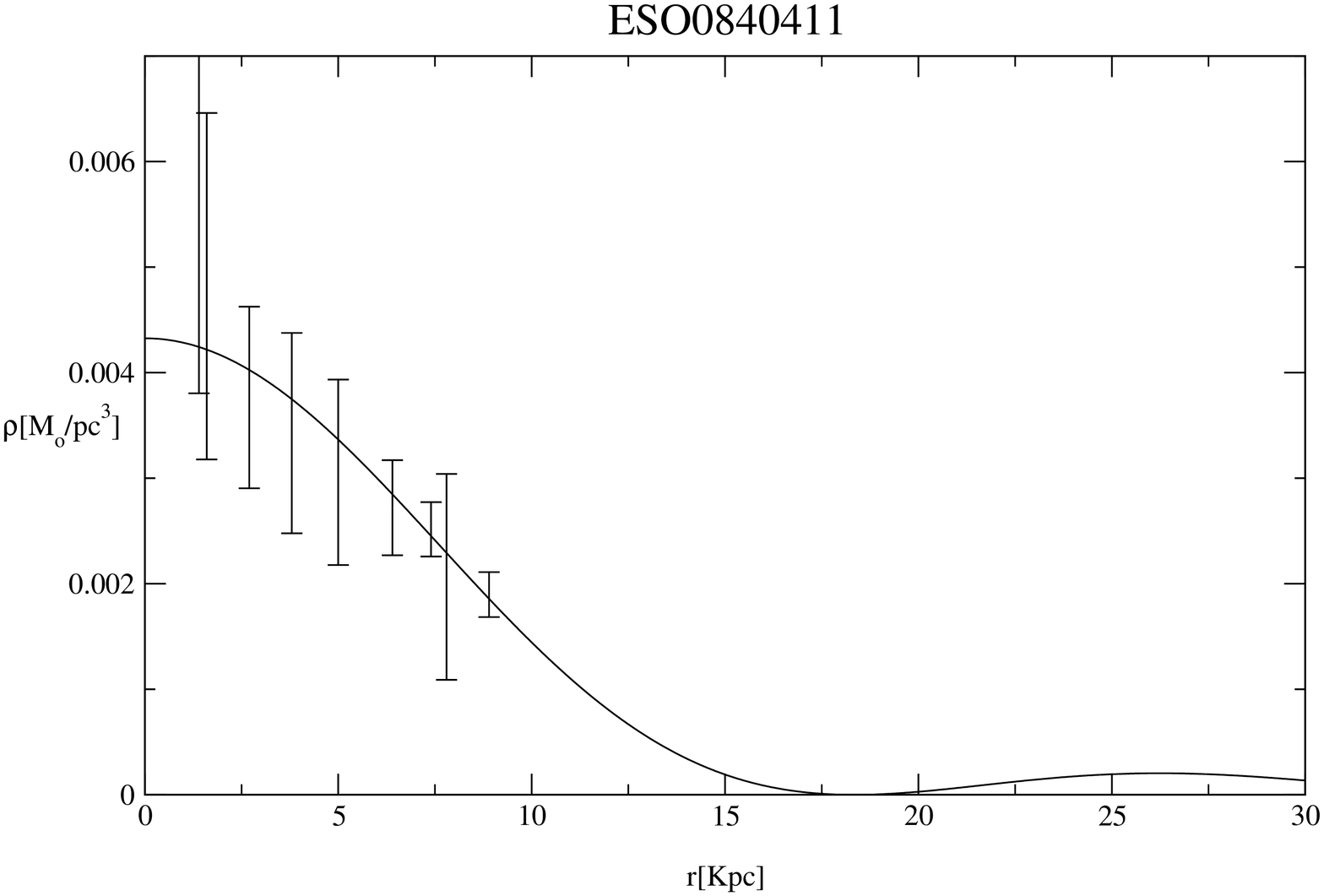}
\caption{\label{fig:G04} Density profile fits for the galaxies
ESO1870510, ESO0488049, ESO0840411. }
\end{center}
\end{figure}

With the scalar field mass $m$ fixed, the profiles fits were made
for the $\Omega$ and $\phi_0$ values with good $\chi^2$
statistic, see Table ~\ref{Table1}. In most of the cases the non
central observational data were the good fitted points, those
data points also have smaller error bars. The density profile fits are
in Figures
\ref{fig:G01},\ref{fig:G02},\ref{fig:G03},\ref{fig:G04}, where we
show several galaxies with the density computed from the observed
rotational profile versus the density obtained with our SFDM
description. In some of them we were able to compare regions
within less than $0.5$ Kpc.

In Table~\ref{Table1} the fundamental frequency
$\Omega$ is listed for each galaxy. We found that the temporal dependence
for the energy-momentum density profile is harmonic with a
temporal period $T=\pi/\Omega$. The column $\Delta\rho(0)$
corresponds to the maximum change in the central density for a
period of time $T$. Finally for all the galaxies the $\phi_0$
value is well inside of the weak gravitation field limit
$\phi_0\lesssim 10^{-3}$.
\begin{table*}[tbp]
  \newcommand{\DS}{\hspace{6\tabcolsep}} %% Expanded Space between
  %% some cols
%  \setlength{\tabnotewidth}{0.9\textwidth}
%  \setlength{\tabcolsep}{1.33\tabcolsep}
%  \tablecols{7}
 \caption{Galactic Parameter Values} \label{Table1}
\begin{tabular}{c c c c c c c c}
    \toprule
\hline Galaxy     &$\Omega$  & $\phi_0$  & $\chi^2$ &
$\rho(r=0)$&$\Delta\rho(0)$
&      $T$\\
           &          &           &          &$[M_{\odot}/pcs^3]$
&$[M_{\odot}/pcs^3]$ &$[yrs]$\\
\hline ESO0140040 & $1+8\,\,\,10^{-9}$ & $1.87\,\,10^{-3}$ &
$12.366$ & $0.569\,\,10^{-1}$ & $0.755\,\,10^{-8}$
&  $1.603382750\,\,10^7$\\
ESO0840411 & $1+6\,\,10^{-9}$ & $5.95\,\,10^{-4}$ & $1.338 $ &
$0.433\,\,10^{-2}$ & $0.147\,\,10^{-9}$
&  $1.603953416\,\,10^7$ \\
ESO1200211 & $1+53\,\,10^{-9}$& $2.04\,\,10^{-4}$ & $10.062$ &
$0.448\,\,10^{-2}$ & $0.530\,\,10^{-9}$
&  $1.603382679\,\,10^7$\\
ESO1870510 &$1+12\,\,10^{-9}$ & $3.28\,\,10^{-4}$ & $3.190 $ &
$0.265\,\,10^{-1}$ & $0.699\,\,10^{-8}$
&  $1.603382570\,\,10^7$\\
ESO2060140 & $1+18\,\,10^{-9}$& $9.18\,\,10^{-4}$ & $65.421$ &
$0.308\,\,10^{-1}$ & $0.206\,\,10^{-8}$
&  $1.603382735\,\,10^7$\\
ESO3020120 & $1+29\,\,10^{-9}$& $6.88\,\,10^{-4}$ & $16.099$ &
$0.279\,\,10^{-1}$ & $0.170\,\,10^{-8}$
&  $1.603382718\,\,10^7$\\
ESO3050090 & $1+40\,\,10^{-9}$& $4.72\,\,10^{-4}$ & $1.224$  &
$0.181\,\,10^{-1}$ & $0.243\,\,10^{-8}$
&  $1.603382699\,\,10^7$\\
ESO4250180 & $1+4\,\,10^{-9}$ & $1.27\,\,10^{-3}$ & $5.221$  &
$0.132\,\,10^{-1}$ & $0.105\,\,10^{-8}$
&  $1.603382757\,\,10^7$\\
ESO4880049 & $1+3\,\,10^{-9}$ & $7.86\,\,10^{-4}$ & $11.410$ &
$0.377\,\,10^{-1}$ & $0.212\,\,10^{-8}$
&  $1.603382715\,\,10^7$\\
\hline
\end{tabular}
\end{table*}

\section{Conclusions and Future Prospects}
\label{sect:conc}

We have found analytic solutions for the EKG equations, for the
case when the scalar field is consider as a test field in a
Minkowski background, and in the relativistic weak gravitational
field limit at first order in the metric perturbations. With these
solutions we have shown that non-trivial local behavior of the
scalar field holds the collapse of an object formed from scalar
field matter. The scalar field contains non trivial, natural
effective pressures which stop the collapse and prevent the
centers of these objects from having cusp-lke density profiles. Even
within this simple approximation it has been possible to fit, with
relative success, the density profiles for some galaxies showing
non cuspy profiles.

Together, all the features of the SFDM model allow one to consider
this model as a robust alternative candidate to be the dark matter
of the Universe, as was suggested by Guzm\'an and Matos 2000,
Matos and Guzm\'an 2000, 2001, and Matos et al 2000. Furthermore,
it has been shown previously that dark halos of galaxies could be
scalar solitonic objects, even in the presence of baryonic matter,
Hu et al 2000; Lee and Koh 1996; Arbey et al 2001, 2002; Sin 1994;
and Ji and Sin 1994. Actually, the boson mass estimated in all
these different approaches roughly coincides with the value
$m_{\Phi }\sim 10^{-23}eV$, even if the later was estimated from a
cosmological point of view, Matos and Ure\~{n}a-L\'{o}pez 2001. We
can appreciate the non-trivial characteristics of the proposed
potential (\ref{coshpot}): Its strong self-interaction provides a
reliable cosmological scenario, while at the same time it has the desired
properties of a quadratic potential. Finally, the
results presented here fill the gap between the successes at
cosmological and galactic levels.

\acknowledgements We would like to thank Miguel Alcubierre,
Vladimir Avila Reese, Arturo Ure\~na and F. Siddhartha Guzm\'an
for many helpful and useful discussions and Erasmo G\'omez and
Aurelio Esp\'{\i}ritu for technical support. We thank Lidia Rojas
for carefull reading and suggestions to improve our work. The
numeric computations were carried out in the "Laboratorio de
Super-C\'omputo Astrof\'{\i}sico (LaSumA) del Cinvestav". This
work was partly supported by CONACyT M\'exico, under grants
32138-E,47209-F and 42748 and by grant number I0101/131/07
C-234/07, Instituto Avanzado de Cosmologia (IAC) collaboration.
Also by the bilateral Mexican-German project DFG-CONACyT 444
MEX-13/17/0-1.

\end{document}